\documentclass[11pt,a4paper]{elsarticle}
\pdfoutput=1 

\usepackage{moreverb}

\usepackage{mathptmx}

\usepackage[colorlinks,bookmarksopen,bookmarksnumbered,citecolor=red,urlcolor=red]{hyperref}
\usepackage{graphicx,epsf,parskip,times,amsmath,lscape,multirow,threeparttable,pgfplots}
\usepackage{cancel,caption,bm,array,setspace,tcolorbox}
\usepackage{subcaption}
\usepackage{floatrow}

\usepackage{etoolbox}
\makeatletter
\appto{\ps@pprintTitle}{\renewcommand{\@oddfoot}{}}
\makeatother

\newcommand\BibTeX{{\rmfamily B\kern-.05em \textsc{i\kern-.025em b}\kern-.08em
T\kern-.1667em\lower.7ex\hbox{E}\kern-.125emX}}

\usepackage[margin=2.5cm]{geometry}

\begin{document}
\begin{frontmatter}
\title{Multiple imputation in Cox regression when there are time-varying effects of exposures}

\author[add1]{Ruth~H.~Keogh\corref{cor1}}
\ead{ruth.keogh@lshtm.ac.uk}
\author[add2]{Tim~P.~Morris}
\ead{tim.morris@ucl.ac.uk}

\cortext[cor1]{Corresponding author}

\address[add1]{Department of Medical Statistics, London School of Hygiene and Tropical Medicine, Keppel Street, London WC1E 7HT, UK.}
\address[add2]{London Hub for Trials Methodology Research, MRC Clinical Trials Unit at UCL, Aviation House, 125 Kingsway, London WC2B 6NH, UK.}

\begin{abstract}
	In Cox regression it is sometimes of interest to study time-varying effects (TVE) of exposures and to test the proportional hazards assumption. TVEs can be investigated with log hazard ratios modelled as a function of time. Missing data on exposures are common and multiple imputation (MI) is a popular approach to handling this, to avoid the potential bias and loss of efficiency resulting from a `complete-case' analysis. Two MI methods have been proposed for when the substantive model is a Cox proportional hazards regression: an approximate method (White and Royston, Statist.\ Med.\ 2009;~28:1982--98) and a substantive-model-compatible method (Bartlett et al., SMMR 2015;~24:462--87). At present, neither method accommodates TVEs of exposures. We extend them to do so for a general form for the TVEs and give specific details for TVEs modelled using restricted cubic splines. Simulation studies assess the performance of the methods under several underlying shapes for TVEs. Our proposed methods give approximately unbiased TVE estimates for binary exposures with missing data, but for continuous exposures the substantive-model-compatible method performs better. The methods also give approximately correct type I errors in the test for proportional hazards when there is no TVE, and gain power to detect TVEs relative to complete-case analysis. Ignoring TVEs at the imputation stage results in biased TVE estimates, incorrect type I errors and substantial loss of power in detecting TVEs. We also propose a multivariable TVE model selection algorithm. The methods are illustrated using data from the Rotterdam Breast Cancer Study. Example R code is provided.
\end{abstract}
\end{frontmatter}

\section{Introduction}
\label{sec:intro}

The setting of this paper is studies of associations between exposures and time-to-event outcomes, such as disease diagnosis or death, analysed using Cox regression \cite{Cox:1972,Cox:1975}. Missing data in explanatory variables are common and the impact of ignoring the missing data and performing a `complete-case' analysis on the subset of individuals with no missing data are loss of efficiency and, depending on the missing data mechanism, biased estimates. Multiple imputation (MI) is a widely used approach for handling missing data that involves generating multiple plausible values for the missing data to create multiple imputed datasets. The multiply imputed datasets are each analysed to obtain estimates of interest and corresponding standard errors, which are then combined using rules developed by Rubin (1987) \cite{Rubin:1987}. The way in which the plausible values for missing data are obtained is important, and in general use of a mis-specified imputation model results in invalid inferences. In general it is desirable that the imputation model is compatible with the chosen substantive model. There exist a range of methods for performing MI covering different substantive model types  -- see Carpenter and Kenward (2013) \cite{Carpenter:2013} for an overview. Two MI approaches have been described for imputation of missing data on covariates in Cox regression. White and Royston (2009) \cite{White:2009} outlined an approximately compatible method which can be implemented in standard software, and Bartlett et al. (2015) \cite{Bartlett:2015} described an alternative `substantive model compatible' approach which does not require approximations.

In time-to-event analyses it is often of interest to study whether the association of certain covariates with the hazard changes over time. Furthermore, assessment of whether the covariate effect changes over time is the basis of a test of the proportional hazards assumption, which is important aspect of model assessment in Cox regression. Ignoring time-varying effects (TVE) and estimating an `average' hazard ratio can result in misleading conclusions \cite{Schemper:1992}.  Cox \cite{Cox:1972} described an extended version of his model to incorporate time-varying effects (TVE) of covariates and there is a large literature on methods for estimating and testing for TVEs in Cox regression: Therneau and Grambsch (2013) \cite{Therneau:2013} (Chapter 6) summarise some of the more popular methods. There is also a more recent literature on model building in Cox regression incorporating TVEs \cite{Buchholz:2011,Abrahamowicz:2007,Wynant:2014,Sauerbrei:2007,Binquet:2008}. 

The existing imputation methods for handling missing data in Cox regression \cite{White:2009,Bartlett:2015} do not account for TVEs of covariates, which could result in invalid inferences. In this paper we extend the methods of White and Royston (2009) \cite{White:2009} and Bartlett et al. (2015) \cite{Bartlett:2015} to accommodate imputation of covariates modelled with TVEs in the Cox regression model. The methods are presented for a general form for a TVE. Specific details are given for TVEs modelled using restricted cubic splines, which are flexible and do not require a form for the TVE to be pre-specified. We also present a model selection algorithm which incorporates imputation of missing data into a procedure for testing for proportional hazards, and  selecting a flexible functional form for TVEs. Throughout, we make the assumption that data are `missing at random' (MAR) \cite{Rubin:1987,Seaman:2013}. Although the term `time-varying effect' is used, we note that a hazard ratio changing over time does not necessarily correspond to a covariate's causal effect changing over time, but may instead occur when the association between a baseline covariate and the hazard becomes weaker (for example) over time, or due to time-varying confounding.

The paper is organised as follows. In Section \ref{sec:general.MI} we describe extensions to the methods of White and Royston (2009) \cite{White:2009} and Bartlett et al. (2015) \cite{Bartlett:2015} to accommodate TVEs, for a general functional form for the TVEs. Use of restricted cubic splines to model the TVEs is described in Section \ref{sec:func.form}. In Section \ref{sec:model.selection} we discuss testing the proportional hazards assumption and present a model selection algorithm. The proposed methods are investigated using simulation studies, described in Section \ref{sec:sim}, in which several underlying functional forms for the TVEs are considered. The methods are illustrated using data from the Rotterdam Breast Cancer Study in Section \ref{sec:example} and we conclude with a discussion in Section \ref{sec:discussion}. Supplementary Materials provide additional details and R code for implementation of the methods.

\section{MI for Cox regression with time-varying effects (TVE)}
\label{sec:general.MI}

\subsection{Preliminaries}

Let $T$ denote an event or censoring time and $D$ be an indicator of whether an individual had the event ($D=1$) or was censored ($D=0$). For simplicity we focus on a single covariate $X_{1}$ with missing data and a fully observed covariate, $X_{2}$. Extensions to missingness in several variables are described in the Supplementary Materials (Section S2). Under the extended Cox model that allows TVEs of covariates \cite{Cox:1972,Therneau:2013}, the hazard function can be written in the general form
\begin{equation}
h(t|X_{1},X_{2})=h_{0}(t)\exp\left\{f_{X_{1}}(t;\bm{\beta}_{X1})X_{1}+f_{X_{2}}(t;\bm{\beta}_{X2})X_{2}\right\}
\label{eq:gen.haz}
\end{equation}
where $h_{0}(t)$ is the baseline hazard and the potential TVEs for $X_{1}$ and $X_{2}$ are represented respectively by the functions $f_{X_{1}}(t;\bm{\beta}_{X1})$ and $f_{X_{2}}(t;\bm{\beta}_{X2})$. Under the standard Cox proportional hazards model, i.e. with no TVEs, $f_{X_{1}}(t;\bm{\beta}_{X1})=\beta_{X1}$ and $f_{X_{2}}(t;\bm{\beta}_{X2})=\beta_{X2}$.

\subsection{MI overview}

Using MI, the general procedure for obtaining estimates of the model parameters $\bm{\beta}_{X1}$ and $\bm{\beta}_{X2}$ is as follows (\cite{Carpenter:2013}, p. 39).
A model $p(X_{1}|T,D,X_{2}; \alpha)$ with non-informative prior on parameters $\alpha$ is specified for $p(X_{1}|T,D,X_{2})$, the distribution of $X_{1}$ given $T$, $D$ and $X_{2}$. Then, for $m=1, \ldots, M$,
\begin{enumerate}
\item 
A value $\alpha^{(m)}$ is drawn from its posterior distribution given the observed data.
\item
For each individual $i$ with missing $X_{1i}$, a value $X_{1i}^{(m)}$ is drawn from $p(X_{1i} | T_i, D_i, X_{2i}; \alpha^{(m)})$, giving an `imputed' data set in which there are no missing values.
\item The substantive model, here the Cox regression model, is fitted to this imputed data set to give estimates $(\hat{\bm{\beta}}^{(m)}_{X1}, \hat{\bm{\beta}}^{(m)}_{X2})$ of $(\bm{\beta}_{X1}, \bm{\beta}_{X2})$, and a corresponding estimate $\hat{\Sigma}^{(m)}$ of ${\rm Var} (\hat{\bm{\beta}}^{(m)}_{X1}, \hat{\bm{\beta}}^{(m)}_{X2})$.
\end{enumerate}
Estimates $(\hat{\bm{\beta}}^{(m)}_{X1}, \hat{\bm{\beta}}^{(m)}_{X2})$ ($m=1, \ldots, M$) and $\hat{\Sigma}^{(m)}$ are combined using `Rubin's rules' \cite{Rubin:1987} to give an overall estimate of $(\bm{\beta}_{X1} ,\bm{\beta}_{X2})$ and of ${\rm Var} (\bm{\beta}_{X1} ,\bm{\beta}_{X2})$.

The main difficulty which arises when the substantive model is a Cox regression is that $p(X_{1}|T,D,X_{2})$ is a non-standard distribution which is only semi-parametrized since $h_{0}(t)$ is non-parametric; therefore we cannot easily draw values from the distribution $p(X_{1}|T,D,X_{2})$ to obtain the imputations. Although in principle any model $p(X_{1}|T,D,X_{2}; \alpha)$ could be used, potentially serious (asymptotic) bias in the estimators of $(\bm{\beta}_{X1} ,\bm{\beta}_{X2})$ and ${\rm Var} (\bm{\beta}_{X1} ,\bm{\beta}_{X2})$ could arise if this model is misspecified. In particular, assuming the substantive model is correctly specified, if the imputation model is not compatible with the substantive model, under certain conditions this implies the imputation model is misspecified \cite{Bartlett:2015}. Consequently, it is desirable that these two models be compatible (or approximately compatible), i.e.\ that there exists a model for the joint distribution $(X_{1}, T, D | X_{2})$ that implies as submodels the model $p(X_{1}|T,D,X_{2}; \alpha)$ used for imputation and the Cox model used for analysis. Two different approaches to using a compatible, or approximately compatible, imputation model have been described by White and Royston (2009) \cite{White:2009} and Bartlett et al. (2015) \cite{Bartlett:2015}, which we refer to respectively as MI-Approx and MI-SMC. In the next two subsections we describe extensions of these imputation methods to accommodate TVEs in the Cox regression model. 

\subsection{MI-TVE-Approx}

For the standard Cox proportional hazards model assuming no TVEs ($f_{X1}(t;\bm{\beta}_{X1})=\beta_{X1}$ and $f_{X2}(t;\bm{\beta}_{X2})=\beta_{X2}$), White and Royston (2009) \cite{White:2009} showed that an approximately compatible imputation model for $X_{1}$ is a logistic regression (for binary $X_{1}$) or linear regression (for continuous $X_{1}$) with linear predictor including main effects of $D$, $X_{2}$, $\widehat{H}(t)$ and the interaction between  $X_{2}$ and $\widehat{H}(t)$, where $\widehat{H}(t)$ is the Nelson-Aalen estimate of the cumulative hazard. Investigations have found, in the settings examined, that the interaction term adds little \cite{White:2009,Borgan:2015}.

When the substantive model is the extended Cox model with TVEs in \ref{eq:gen.haz}, we can show that an approximately compatible imputation model for $X_{1}$ is a logistic regression (for binary $X_{1}$) or linear regression (for continuous $X_{1}$) with linear predictor including main effects of $X_{2}$, $D f_{X1}(T)$, $\widehat{H}(T)$, $\widehat{H}^{(1)}(T)$ and interactions of $X_{2}$ with $\widehat{H}(T)$ and $\widehat{H}^{(1)}(T)$, where $\widehat{H}^{(1)}(T)$ is the Nelson-Aalen-type estimator $\widehat{H}^{(1)}(T)=\sum_{t\leq T}\frac{td(t)}{n(t)}$ ($d(t)$ and $n(t)$ denote the number of events and number at risk at time $t$). The details of the derivation are given in the Supplementary Materials (Section S1).  We refer to the resulting approach as MI-TVE-Approx. In the simulations we will investigate whether it is important to include the higher order cumulative hazard term $\widehat{H}^{(1)}(T)$ and/or the interaction terms $X_{2}\times \widehat{H}(T)$ and $X_{2}\times \widehat{H}^{(1)}(T)$. The imputation procedure is as follows. 

\begin{enumerate}
	\item Fit the imputation model to the subset of individuals with complete data. For continuous $X_{1}$ this is
    {\small
$$
    X=\alpha_{0}+\alpha_{1}X_{2}+\alpha_{2}^{\prime}D f_{X1}(T)+\alpha_{3}\widehat{H}(T)+\alpha_{4}\widehat{H}^{(1)}(T)+\alpha_{5}X_{2}\widehat{H}(T)+\alpha_{6}X_{2}\widehat{H}^{(1)}(T)+\epsilon,
    \label{eq:imp.model.step1.linear}
$$} 
    and for binary $X_{2}$
    {\small
$$
		\mbox{logit }\mbox{Pr}(X_{1}=1|T,D,X_{2})=\alpha_{0}+\alpha_{1}X_{2}+\alpha_{2}^{\prime}D f_{X1}(T)+\alpha_{3}\widehat{H}(T)+\alpha_{4}\widehat{H}^{(1)}(T)+\alpha_{5}X_{2}\widehat{H}(T)+\alpha_{6}X_{2}\widehat{H}^{(1)}(T).
		\label{eq:imp.model.step1/logistic}
$$}
	\item Take $M$ random draws values of the parameters from their approximate posterior distribution (we refer to \cite{White:2011} for details), denoted $\alpha^{(m)}_{j}$ $(j=0,1\ldots,6)$ (binary and continuous $X_{1}$, and additionally $\sigma^{2(m)}_{\epsilon}$ for continuous $X_{1}$
	\item The imputed value of $X_{1i}$ in the $m$th imputed data set is given (for continuous $X_{1}$) by $X_{1i}^{(m)}=\alpha^{(m)}_{0}+\alpha^{(m)}_{1}X_{2}+\alpha^{(m)\prime}_{2}D f_{X1}(T)+\alpha^{(m)}_{3}\widehat{H}(T)+\alpha^{(m)}_{4}\widehat{H}^{(1)}(T)+\alpha^{(m)}_{5}X_{2}\widehat{H}(T)+\alpha^{(m)}_{6}X_{2}\widehat{H}^{(1)}(T)+\epsilon^{(m)}.$, where $\epsilon^{(m)}$ is a random draw from a normal distribution with mean 0 and variance $\sigma^{2(m)}_{\epsilon}$. For binary $X_{1}$, the imputed value is a draw from a Bernoulli distribution with $\mbox{logit } \mbox{Pr}(X_{1}=1|T,D,X_{2})=\alpha^{(m)}_{0}+\alpha^{(m)}_{1}X_{2}+\alpha^{(m)\prime}_{2}D f_{X1}(T)+\alpha^{(m)}_{3}\widehat{H}(T)+\alpha^{(m)}_{4}\widehat{H}^{(1)}(T)+\alpha^{(m)}_{5}X_{2}\widehat{H}(T)+\alpha^{(m)}_{6}X_{2}\widehat{H}^{(1)}(T)$.
\end{enumerate}

\subsection{MI-TVE-SMC}

In the context of the standard Cox proportional hazards model without TVEs, MI-Approx has been found to work well in a range of circumstances \cite{White:2009,Carpenter:2013}. However, the approximation can perform badly in some `extreme' situations, including when there are large effect sizes and when the event rate is high \cite{White:2009}. Bartlett et al (2015) \cite{Bartlett:2015} described an approach, referred to here as MI-SMC, which ensures the imputation model is compatible with the user's chosen substantive model, here a Cox regression (`substantive model compatible' -- SMC). However, they did not accommodate TVEs. We outline this extension, which was first described by Bartlett \cite{Bartlett:2010} in the context of time-dependent covariates, and refer to the resulting method as MI-TVE-SMC. The MI-TVE-SMC imputation procedure is as follows. 

First, a model $p(X_{1}|X_{2};\gamma_{X1})$ is specified. For binary $X_{1}$ this may be a logistic regression model and for continuous $X_{1}$ a linear regression model. The steps used to obtain the $m$th imputed data set are then:
\begin{enumerate}
\item Fill in the missing variables with arbitrary starting values, to create a complete data set.
\item Fit the Cox regression model of interest, including TVEs, to the current complete data set to obtain estimates $(\bm{\hat{\beta}}_{X1},\bm{\hat{\beta}}_{X2})$ and their estimated variance $\widehat{\Sigma}$.
Draw values $\bm{\beta}^{(m)}_{X1},\bm{\beta}^{(m)}_{X2}$ from a joint normal distribution with mean $(\hat{\beta}_{X1},\hat{\beta}_{X2})$ and variance $\widehat{\Sigma}$.
\item Calculate Breslow's estimate \cite{Breslow:1972}, denoted $H^{(m)}_{0}(t)$, of the baseline cumulative hazard $H_{0}(t)$ using parameter values $\bm{\beta}^{(m)}_{X1},\bm{\beta}^{(m)}_{X2}$ and current imputations of $X_{1}$.
\item Estimate parameters $\gamma_{X1}$ and their variance by fitting the assumed regression model for $X_{1}$ on $X_{2}$ to the current complete data set. Draw a value $\gamma_{X1}^{*}$ from the approximate joint posterior distribution of $\gamma_{X1}$ \cite{White:2011}.
\item For each individual for whom $X_{1}$ is missing, (a) draw a value $X_{1}^{*}$ from the distribution $p(X_{1}|X_{2};\gamma_{X1}^{*})$, and (b) draw a value $U$ from a uniform distribution on $[0,1]$. Accept the value $X_{1}^{*}$ if
{\footnotesize
$$
\left\{
\begin{array}{ll}
U\leq \exp \left[- \sum_{j:t_{j} \leq T} \Delta H^{(m)}_{0}(t_{j}) \exp\left\{f_{X1}\left(t_{j};\bm{\beta}^{(m)}_{X1}\right)X_{1}^{*} +f_{X2}\left(t_{j};\bm{\beta}^{(m)}_{X2}\right) X_{2}\right\}  \right]&\mbox{if }D=0\\
U\leq \Delta H^{(m)}_{0}(T)\exp\left\{1+f_{X1}\left(T;\bm{\beta}^{(m)}_{X1}\right)X_{1}^{*} +f_{X2}\left(T;\bm{\beta}^{(m)}_{X2}\right) X_{2}- \sum_{j:t_{j} \leq T} \Delta H^{(m)}_{0}(t_{j})e^{f_{X1}\left(t_{j};\bm{\beta}^{(m)}_{X1}\right)X_{1}^{*} +f_{X2}\left(t_{j};\bm{\beta}^{(m)}_{X2}\right) X_{2}}\right\}&\mbox{if }D=1
\end{array}\right.
$$}
where $\Delta H^{(m)}_{0}(t)$ denotes the increment in $H^{(m)}_{0}(t)$ at time $t$ and $t_{1},..,t_{k}$ denote the unique failure times. Repeat (a) and (b) until a value $X_{1}^{*}$ is accepted.
\item Return to steps 2--5 until the imputed $X_{1}$ values have converged to a stationary distribution. These are then the imputed values in the $m$th imputed data set.
\end{enumerate}
The difference between the MI-SMC approach, which does not accommodate TVEs, and the MI-TVE-SMC approach is in the terms used for the rejection in step 5, and the fact that a Cox model with TVEs is fitted in Step 2.

We have outlined the MI-TVE-Approx and MI-TVE-SMC approaches for the simple setting of missing data in a single covariate $X_{1}$ with a TVE. Both methods extend to handle missingness in several covariates using the fully conditional specification (FCS) approach (also referred to as multiple imputation by chained equations), in which an imputation model is specified for each partially missing covariate conditional on all the other covariates and an iterative approach is used to fit the imputation models \cite{vanBuuren:2007,White:2011}. Details are provided in the Supplementary Materials (Section S2). 

\section{Functional form of time-varying effects (TVE)}
\label{sec:func.form}

In the preceding section the MI methods were described for a general functional form for the TVEs, $f_{X}(t;\bm{\beta}_{X})$. Approaches to modelling TVEs include use of pre-specified parametric functional forms \cite{Quantin:1999} (e.g. $f_{X}(t;\bm{\beta}_{X})=\beta_{0}+\beta_{1}t$), step-functions \cite{Gore:1984,Moreau:1985,Quantin:1999}, fractional polynomials \cite{Sauerbrei:2007}, and splines \cite{Hess:1994,Heinzl:1996,Abrahamowicz:1996,Abrahamowicz:2007,Wynant:2014,Hastie:1993,Gray:1992,Kooperberg1995}. In this paper we focus on using a restricted cubic spline form \cite{Hess:1994, Heinzl:1996} for the TVE function because they allow a flexible form for the TVE with relatively few parameters. Under a restricted cubic spline with $L$ knots at $u_{1},\ldots,u_{L}$ the TVE function for a covariate $X$ is
{\small
	\begin{equation}
	f_{X}(t;\bm{\beta}_{X})=\beta_{X0}+\beta_{X1}t+\sum_{i=1}^{L-2}\theta_{Xi}\left\{(t-u_{i})^{3}_{{\tiny +}}-\left(\frac{(t-u_{L-1})^{3}_{{\tiny +}}(u_{L}-u_{i})}{(u_{L}-u_{L-1})}\right)+\left(\frac{(t-u_{L})^{3}_{{\tiny +}}(u_{L-1}-u_{i})}{(u_{L}-u_{L-1})}\right)\right\}
    \label{eq:rcs}
\end{equation}}
where $(t-u_{i})_{{\tiny +}}$ takes value $(t-u_{i})$ if $(t-u_{i})>0$ and 0 otherwise. 
The number of knots used, and the position of the knots, has to be decided by the user and there is no formal theoretical basis for the decision. Hess (1994) \cite{Hess:1994} noted empirical evidence that three to five knots are usually adequate and the fit is not greatly altered by altering the knot positions. Stone (1986) \cite{Stone:1986} also recommended using 5 knots in restricted cubic splines. Hess (1994) \cite{Hess:1994} suggested placing knots at quantiles of the observed follow-up times (including both event times and censoring times); including the outer knots near the extremes; and placing the knots approximately uniformly over the quantiles of the distribution of the follow-up times. Similar recommendations were given by Durrleman and Simon (1989) \cite{Durrleman:1989} in the context of restricted cubic splines for functional forms of covariates in survival analyses. In the simulations we consider using restricted cubic splines with 5 knots placed at percentiles (5th, 25th, 50th, 75th, 95th) of the event time distribution. 

In the case of a restricted cubic spline with $L=5$ knots, the MI-TVE-Approx imputation model for $X_{1}$ should include $X_{2}$, $D$, the interaction between $D$ and $T$, interactions of $D$ with \\$\left\{(T-u_{i})^{3}_{{\tiny +}}-\left(\frac{(T-u_{L-1})^{3}_{{\tiny +}}(u_{L}-u_{i})}{(u_{L}-u_{L-1})}\right)+\left(\frac{(T-u_{L})^{3}_{{\tiny +}}(u_{L-1}-u_{i})}{(u_{L}-u_{L-1})}\right)\right\}$ (for $i=1,2,3$), $\widehat{H}(T)$, $\widehat{H}^{(1)}(T)$ and interactions of $X_{2}$ with $\widehat{H}(T)$ and $\widehat{H}^{(1)}(T)$.

\section{Testing the proportional hazards assumption and model selection}
\label{sec:model.selection}

In most contexts, when using Cox regression modelling it is important to assess whether covariates have TVEs, that is, to perform tests of the proportional hazards assumption. TVEs can then be included for covariates for which the proportional-hazards assumption appears not to hold. Tests of proportional hazards based on TVEs modelled using splines have been previously described by Abrahamowicz et al.\ \cite{Abrahamowicz:1996}. With fully observed data on covariates, the proportional hazards assumption can be tested using a likelihood ratio test, comparing a model including TVEs to the model without TVEs. A joint Wald test of the TVE parameters is asymptotically equivalent: assuming a restricted cubic spline for the TVE for $X$ (equation \ref{eq:rcs}, this is a joint test of $\beta_{X1}=\theta_{X1}=\ldots =\theta_{X,L-1}=0$. Wood et al.\ \cite{Wood:2008} described the use of Wald tests for model selection using multiply imputed data, and this was further evaluated by Morris et al.\ \cite{Morris:2015} in the context of covariate transformations based on fractional polynomials. We suggest this approach for tests of TVEs. The joint Wald test of the parameters of interest (null hypothesis $\beta_{X1}=\theta_{X1}=\ldots =\theta_{X,L-1}=0$) is performed using the parameter estimates and the corresponding variance covariance matrix obtained from Rubin's rules. For the purposes of testing the proportional hazards assumption as part of a model assessment and selection procedure, we recommend allowing TVEs for all variables at the imputation stage of the analysis; the importance of doing so for valid tests of the proportional hazards assumption is investigated in the simulation studies.

Finally, we propose an algorithm (the MI-MTVE algorithm) which provides a model selection procedure for identifying TVEs using multiply imputed data. Several authors have proposed algorithms for model selection involving both TVEs and transformation of covariates \cite{Abrahamowicz:2007,Wynant:2014,Sauerbrei:2007}, though all assume fully observed datasets. The MI-MTVE algorithm is an adaptation of the MFPT algorithm of Sauerbrei et al.\ \cite{Sauerbrei:2007}, which uses fractional polynomial transformations of covariates and fractional polynomial forms for TVEs. Our adaptation employs restricted cubic spline transformations, rather than fractional polynomials, for TVEs, and is similar to a procedure advocated by Wynant and Abrahamowicz (2014) \cite{Wynant:2014}. Forwards selection is used to accommodate investigation of TVEs in multiple covariates and selection of a functional form for TVEs using restricted cubic splines with up to 5 knots.

\textbf{MI-MTVE algorithm}
\begin{description}
  \item[Step 1] Perform MI-TVE-Approx or MI-TVE-SMC, assuming a restricted cubic spline with 5 knots for the TVE for each covariate, to obtain $M$ imputed data sets.
  \item[Step 2] In each imputed data set, fit the model with no TVEs of any covariate (denoted model $\mathcal{M}_0$). Denote the set of covariates by $C$. For each $c~(c \in C)$, fit four TVE models of increasing complexity (indexed by $j$) to each imputed data set: linear form ($f_{X}(t;\bm{\beta}_{X})=\beta_{X0}+\beta_{X1}t$), and restricted cubic splines with 3, 4 and 5 knots. 
  \item[Step 3] For each covariate $c$, test for TVEs in each model $j$ using joint Wald tests of the TVE parameters based on Rubin's rules. Select the combination of covariate ($c$) and TVE model ($j$) which returns the smallest p-value in the test for TVEs. If no combination of $c$ and $j$ gives a p-value less than a chosen level $\alpha$, stop; the working model without TVEs $\mathcal{M}_0$ is adequate. Otherwise, update the working model $\mathcal{M}_0$ to include TVEs for the covariate $c$ and TVE model $j$ which returned the smallest p-value. Call this new working model $\mathcal{M}_1$.
  \item[Step 4] Repeat steps 2--3 with updated working models until there are no remaining covariates $c$  not in the current working model that have a significant TVE (at level $\alpha$) under any TVE model $j$. Stop; this working model is the final selected model.
\end{description}
The estimates of the parameters of the final selected model, and corresponding estimated covariance matrix, are those obtained by applying Rubin's rules to the results from fitting the final model to each imputed data set. The MI is performed only in Step 1 and is based on a TVE for each covariate of the most complex form that we consider in this paper (a restricted cubic spline with 5 knots). This means that a restriction of the imputation model should be compatible with the model selected by the algorithm (termed `semi-compatibility' \cite{liu14,Bartlett:2015,Morris:2015}). The imputation model may include some redundant parameters, but this will not impact on the validity of MI inference.

\section{Simulation study} \label{sec:sim}
We now present a simulation study which was designed to evaluate the MI methods across a variety of data-generating scenarios.

\subsection{Data-generating mechanisms} \label{subs:dgms}
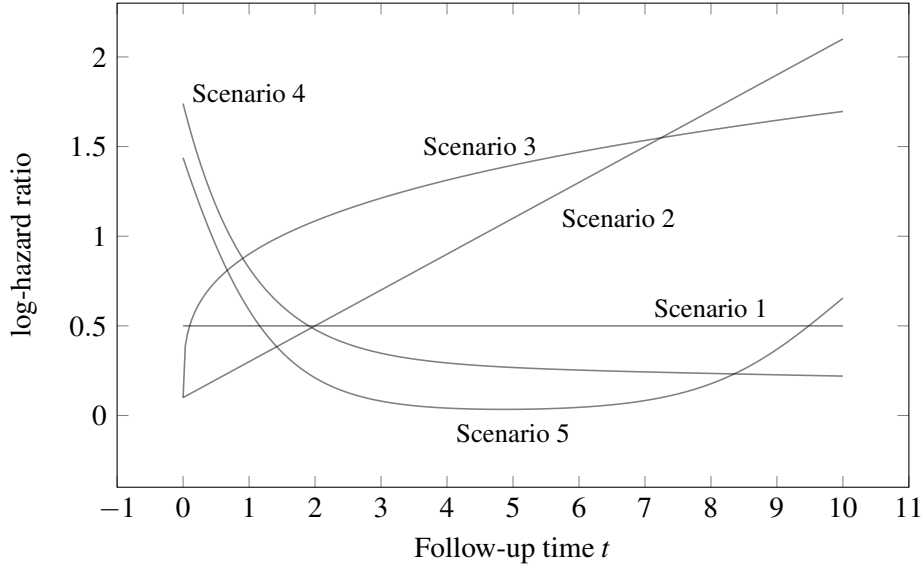
\begin{figure}
\caption{Time varying effect functions used in simulation studies.}
	\label{fig:tves}
\begin{center}
  \begin{tabular}{ccll}
  \hline
  Scenario & TVE & $f_{X_1}(t)$ & $f_{X_2}(t)$ \\
  \hline
  1 & -- & $0.5$ & $0.5$ \\
  2 & $X_1$ & $0.1+0.2t$ & $0.5$ \\
  3 & $X_1$ & $0.1+0.8t^{0.3}$ & $0.5$ \\
  4 & $X_1$ & $0.32+1.42e^{-t}-0.02t^{0.7}$ & $0.5$ \\
  5 & $X_1$ & $\frac{4}{1+e^{1.2(t+0.5)}}+\frac{4}{3(1.1+e^{10-t})}+0.02$ & $0.5$ \\
  \hline
  \end{tabular} \bigskip \\
\begin{tikzpicture}[baseline]
\begin{axis}[
  samples=300,
  domain=0:10,
  height=8cm,
  width=12cm,
  restrict y to domain=-.5:2.1,
  restrict x to domain=0:10,
  xlabel=Follow-up time $t$, ylabel=log-hazard ratio,
  ymin=-.4,ymax=2.3,
  ]
  \addplot[black,semithick,opacity=0.5 ] plot(\x,{.5});
  \addplot[black,semithick,opacity=0.5 ] plot(\x,{.1+.2*\x});
  \addplot[black,semithick,opacity=0.5 ] plot(\x,{.1 + .8*(\x^.3)});
  \addplot[black,semithick,opacity=0.5 ] plot(\x,{.32 + (1.42*exp(-\x)) -(.02*(\x^.7)});
  \addplot[black,semithick,opacity=0.5 ] plot(\x,{(4/(1 + exp(1.2*(\x+0.5)))) + (4/(3*(1.1+exp(10-\x)))) + 0.02)});
  \node[align=center] at(800,100) {\small Scenario 1};
  \node[align=center] at(660,150) {\small Scenario 2};
  \node[align=center] at(450,190) {\small Scenario 3};
  \node[align=center] at(100,220) {\small Scenario 4};
  \node[align=center] at(500,30)  {\small Scenario 5};
\end{axis}
\end{tikzpicture}
\end{center}
\end{figure}

Data were generated for a cohort of 2,000 individuals. Two covariates, $X_{1}$ and $X_{2}$, are considered. Event times $T_{E}$ were generated according to the exponential hazard model
\begin{equation}
	h(t|X_{1},X_{2})=\lambda_{E}\exp\left\{f_{X1}(t;\bm{\beta}_{X1})X_{1}+f_{X2}(t;\bm{\beta}_{X2}) X_{2}\right\}
\label{eq:sim.model}
\end{equation}
We consider five forms for the TVEs. These are listed in the table at the top of Figure \ref{fig:tves}. In scenario 1, neither covariate has a time-varying effect. In scenarios 2--5, $X_1$ has TVE but $X_2$ does not. Figure \ref{fig:tves} depicts the form of the TVEs. These forms include examples previously used by Buchholz \cite{Buchholz:2010} and Buchholz and Sauerbrei \cite{Buchholz:2011}.

Random drop out times, $T_{C}$, were generated according to an exponential distribution with rate $\lambda_{C}$, and administrative censoring was imposed after 10 years of follow-up. The observed time for each individual was calculated as $T=\mathrm{min}(T_{E},T_{C},10)$. Values for $\lambda_{E}$ and $\lambda_{C}$ were chosen such that 10\% of individuals have the event of interest and 50\% are censored due to random drop out, with the remainder being administratively censored.

Both binary and continuous covariates are considered. Binary $X_{1}$ was generated from a binomial distribution such that $P(X_{1}=1)=0.2$ and binary $X_{2}$ was generated using $\mbox{logit}\{P(X_{2}=1|X_{1})\}=X_{1}$. Continuous $X_{1}$ and $X_{2}$ were generated from a bivariate normal distribution with means 0, variances 1 and correlation 0.5.

Non-monotone missing data were generated in $X_{1}$ and $X_{2}$ according to a MAR mechanism in which the probability of missingness in $X_{1}$ depends on observed values of $X_{2}$, and vice versa (see Supplementary Materials Section S4). In this, $X_{1}$ is missing for 30\% of individuals and $X_{2}$ for 30\% of individuals, resulting in approximately 50\% of individuals missing at least one of the measurements.

There are 10 main simulation scenarios: five different scenarios for TVEs of $X_1$, and binary and continuous $X_{1}$ and $X_{2}$. Five hundred simulated data sets were generated under each scenario (justified in the Supplementary Materials Section S5). In Section \ref{sec:sims.extra} we present results from additional sensitivity scenarios with a higher event rate, lower level of missingness and a MAR mechanism in which the missingness in $X_{1}$ and $X_{2}$ additionally depends on the outcome $D$.



\subsection{Methods compared}

The methods we investigate are the proposed MI-TVE-Approx and MI-TVE-SMC approaches, and for comparison the corresponding approaches which do not incorporate TVEs (MI-Approx and MI-SMC). We also performed a complete-data analysis (before missing data is introduced) and a complete-case analysis, which uses only the subset with no missing data. In MI-Approx we omitted the interaction terms between covariates and  $\widehat{H}(T)$, because their inclusion was not found to result in material differences in the results. For the same reason, in MI-TVE-Approx we omitted the interaction terms and terms involving $\widehat{H}_{1}(T)$. The MI-TVE-Approx imputation model recommended for $X_{1}$ therefore includes $X_{2}$, $D$, the interaction between $D$ and $T$, interactions of $D$ with $\left\{(T-u_{i})^{3}_{{\tiny +}}-\left(\frac{(T-u_{L-1})^{3}_{{\tiny +}}(u_{L}-u_{i})}{(u_{L}-u_{L-1})}\right)+\left(\frac{(T-u_{L})^{3}_{{\tiny +}}(u_{L-1}-u_{i})}{(u_{L}-u_{L-1})}\right)\right\}$ for $i=1,2,3$, and $\widehat{H}(T)$. The recommended imputation model for $X_{2}$ is the same but with $X_{2}$ replaced by $X_{1}$.

In all Cox regression analyses TVE for $X_{1}$ and $X_{2}$ are modelled using a restricted cubic spline with 5 knots placed at percentiles (5th, 25th, 50th, 75th, 95th) of the distribution of the observed event times. This includes a TVE model for $X_{2}$ ($f_{X2}(t;\beta_{X2})$) even though in the data generating process there is no TVE of $X_{2}$. In MI-TVE-Approx and MI-TVE-SMC the TVE was incorporated based on the same functional form. 

In the MI analyses we used 10 imputed data sets. For the analysis of studies in practice, we recommend the rule of thumb suggested by White, Royston and Wood (2011) \cite{White:2011} to set the number of imputations to be approximately the same as the percentage of missing data, with a larger number chosen if numerical reproducibility of estimates is desired.



\subsection{Performance measures} \label{subs:perfmeas}

The performance of methods was assessed in a number of ways, described below. Each assessment was performed for both $X_{1}$ and $X_{2}$.
\begin{itemize}
	\item Curve-wise estimate of the TVE, presented visually over the follow-up time and averaged over simulation runs.
 \item Bias in the estimated curve at 1, 5 and 9 years, and corresponding 95\% Monte Carlo confidence intervals. The bias from the MI methods and the complete-case analysis was calculated relative to the complete-data results, i.e. as the difference between the MI or complete-case estimates and the mean of the complete-data estimates. This was done because the true data generating mechanism is not a restricted cubic spline and therefore we do not necessarily expect to get completely unbiased estimates from the complete-data analysis.
\item Coverage of confidence intervals, estimated at 1, 5 and 9 years, defined as the proportion of simulated data sets for which the true curve lies within the 95\% confidence intervals at time $t$.
	\item Rejection fractions for the test of the proportional hazards assumption. For scenario 1, this corresponds to a type I error rate for the TVEs of both $X_1$ and $X_{2}$. For all other scenarios, this corresponds to power for the TVE of $X_1$ and type I error rate for the TVE of $X_2$. The proportional hazards assumption is assessed using a joint Wald test of the TVE parameters. 
\end{itemize}

We generated 500 estimated data sets under each scenario. Justification for this, referring to Monte Carlo errors in the bias and coverage, are given in the Supplementary Materials (Section S5).



All simulations and analyses were performed using R. The substantive model was fitted using \texttt{coxph} in the \texttt{survival} package. MI-Approx and MI-TVE-Approx were implemented using \texttt{mice} \cite{vanBuuren:2011}, and MI-SMC using \texttt{smcfcs} (\url{https://github.com/jwb133/smcfcs}). We extended the \texttt{smcfcs} code to implement MI-TVE-SMC.  Example code for all methods, and an example simulated data set, are given in the Supplementary Materials (Section S6 and additional files).

\subsection{Simulation results}

\subsubsection{Curve-wise estimates and bias}

Figures 2 and 3 show the curve-wise TVE estimates for covariate $X_{1}$ in the binary and continuous covariates settings. Figures 4 and 5 show the bias in the estimated TVE curves at three time points (1, 5, 9), corresponding to the difference between the mean curves shown in Figures 2 and 3 and the true curve. Similar plots for $X_{2}$, which always has a time-constant effect, are shown in the Supplementary Materials (Figures S1 and S2). 

The complete-data and complete-case analyses give approximately unbiased TVE estimates, except for some bias in the complete-case analysis in the extremes of some curves. As noted earlier, the complete-data analysis could give estimates with some slight bias because the data were not generated under the restricted cubic spline model which is used in the analysis. Note that we expect the complete-case analysis to give an approximately unbiased result because the missingness does not depend on the outcome. The MI methods which accommodate TVEs, MI-TVE-Approx and MI-TVE-SMC, perform similarly for binary $X_{1}$ and give estimated TVE estimates similar to that from the complete-data analysis. However, for continuous $X_{1}$ only MI-TVE-SMC performs well in general, with MI-TVE-Approx giving clearly biased estimates in scenarios 2 and 3 at times where the TVE is quite large. MI-TVE-Approx requires additional approximations for continuous covariates and the approximation does not perform well in these scenarios. Poor performance of MI-Approx in scenarios with continuous covariates and large effect sizes has been found previously in the setting without TVEs \cite{White:2009}. The results show that failing to account for the TVE in the imputation, as in MI-Approx and MI-SMC, results in a biased estimate of the TVE curve. The bias is such that the TVE appears attenuated. 

Tables of coverages of the estimated TVE curves at three time points are shown in Supplementary Tables S1 and S2. The coverages tend to be higher than the nominal 95\% level and many are 100\%, including in the complete-data analyses. Coverage not at the nominal level has been previously observed for spline-based models \cite{Cummins:2001}.

\subsubsection{Tests of the proportional hazards assumption}

Tables 1 and 2 show the percentage of simulations in which the proportional hazards assumption (based on joint Wald tests) was rejected at the 5\% level for $X_{1}$ and $X_{2}$, in the binary and continuous covariates settings. In Scenario 1, where neither $X_1$ nor $X_{2}$ has a TVE, the percentage of simulations in which the null hypothesis of proportional hazards was rejected was close to 5\% in the complete-data and complete-case analyses, indicating approximately correct Type I errors. The Type I error rates from MI-TVE-SMC were slightly inflated in some scenarios.   
In scenario 2-5 with TVEs for $X_{1}$, the power to reject the PH null hypothesis varied under the complete-data analysis, from 100\% (continuous covariates, scenario 2)  to 33\% (binary covariates, scenario 3). Power was generally lower in the setting with binary covariates. Power was reduced under the complete-case analysis, for example in scenario 4 with continuous covariates the power from the complete-case analysis was 77\% compared to 99\% in the complete-data analysis, and in scenario 4 with binary covariates the power from the complete-case analysis was 17\% compared to 56\% in the complete-data analysis. For binary covariates the power under MI-TVE-Approx and MI-TVE-SMC was much increased relative to the complete-case analysis and was highest for MI-TVE SMC. Power using MI-TVE-SMC was also high in the setting with continuous covariates, but lower for MI-TVE-Approx, and in scenarios 2 and 3 lower than that from the complete-case analysis; this is partly a consequence of the bias observed using MI-TVE-Approx for continuous covariates. The results show that if the TVEs are ignored in the imputation (MI-Approx and MI-SMC) there is a large loss of power to reject the null hypothesis of proportional hazards across all scenarios, and power from these methods was lower than that from the complete-case analysis. 

\subsection{Additional simulation investigations}
\label{sec:sims.extra}

We investigated the performance of the methods in three additional situations:
\begin{itemize}
\item[(i)] Missingness depends on the outcome (see Supplementary Materials Section S4). Missingness depending on the outcome is plausible if there is an underlying latent feature which is associated with the subsequent outcome and with missingness. 
\item[(ii)] 50\% of individuals have the event. This may not be a common situation but is relevant for certain clinical studies.
\item[(iii)] A lower percentage of individuals with missing covariate data. The percentage of individuals missing $X_{1}$ and missing $X_{2}$ was reduced to 10\% (see Supplementary Materials Section S4), which results in approximately 20\% of individuals missing at least of the measurements. 
\end{itemize}

Other aspects of the simulations were as described above. For additional simulations (i) and (ii) results are presented for scenario 4 (Figure \ref{fig:tves}) with binary covariates, representing a situation in which the association between $X_{1}$ and the hazard becomes weaker over time. For additional simulation (iii) we focused on scenario 2 (Figure \ref{fig:tves}) with continuous covariates, for which we found biased estimates using MI-TVE-Approx in the earlier simulation results.

When the missingness depends on the outcome the complete case analysis gives biased estimates (\ref{fig:sim.MARd}). The results show that the proposed MI methods continue to perform well, while ignoring TVEs in the imputation still results in bias, as we would expect based on our earlier results. The results in Figure \ref{fig:sim.50pcevent} show that the proposed methods continue to perform well in a situation in which 50\% of individuals have the event. When the proportion of individuals with missing data is reduced the bias from MI-TVE-Approx in scenario 2 with continuous covariates is reduced (Figure \ref{fig:sim.10pcmiss}), but still evident when the time-varying effects is large. 

\section{Illustration: Rotterdam Breast Cancer Study} \label{sec:example}

The methods were illustrated using data on 2,982 individuals with primary breast cancer from the Rotterdam tumour bank. This data set is freely available (we used the data set provided at
http://portal.uni-freiburg.de/imbi/Royston-Sauerbrei-book/index.html\#datasets) and was used by Sauerbrei et al (2007) \cite{Sauerbrei:2007} and Royston and Sauerbrei (2008) \cite{Royston:2008} to illustrate time-varying exposure effects. Individuals were followed-up from the time of breast cancer diagnosis to a composite event of the first of disease recurrence or death due to breast cancer. Over the course of follow-up, which ranged from 1 to 231 months, 1,518 individuals (51\%) had the outcome of interest and the remainder were censored. In this illustration we focus on eight variables used by Sauerbrei et al.\ (2007) \cite{Sauerbrei:2007} and Royston and Sauerbrei (2008) \cite{Royston:2008}: age, tumour size 1 ($\leq$ 20mm, $>$ 20mm), tumour size 2 ($\leq$ 50mm, $>$ 50mm), tumour grade (grade 2 or 3 versus grade 1), squared transformed number of positive lymph nodes ($\mathrm{enodes}=\exp(-2\times0.12\times\mathrm{nodes})$), treatment with hormonal therapy (yes vs.\ no), treatment with chemotherapy (yes vs.\ no), and transformed progesterone receptors (pmol/l) ($\log(\mathrm{pgr}+1)$). Sauerbrei et al (2007) \cite{Sauerbrei:2007} and Royston and Sauerbrei (2008) \cite{Royston:2008} detected time-varying effects for two of the variables, tumour size 1 and $\log(\mathrm{pgr}+1)$, via interactions with log time ($f_{X}(t;\bm{\beta}_{X})=\beta_{X0}+\beta_{X1}\log t$ in our notation).

For this illustration we generated missing data at random (MAR) in five variables (tumour grade, $\mathrm{enodes}$, hormonal therapy, chemotherapy, and $\log(\mathrm{pgr}+1)$) with the probability of missingness depending on age and tumour size ($e^{-9+0.1\times \mbox{age}-\mbox{tumour size 2}}/(1+e^{-9+0.1\times \mbox{age}-\mbox{tumour size 2}})$). The missing data were generated conditionally independently for each variable such that approximately 5\% of individuals have missing data in any given variable. This resulted in approximately 20\% of individuals having missing data on at least one variable.

We performed the following analyses: a complete-data analysis before missingness was introduced; a complete-case analysis on the subset with no missing data; MI-Approx; MI-SMC; MI-TVE-Approx; MI-TVE-SMC. The imputations allowing TVEs assumed restricted cubic splines with 5 knots for all covariates. In each analysis the substantive model was fitted first assuming no TVEs. A test of proportional hazards was performed for each covariate in turn (using joint Wald tests), based on TVE models of four forms (linear form, and restricted cubic splines with 3, 4 and 5 knots). The TVE form giving the smallest p-value was selected. This corresponds to the first step of the MI-MTVE algorithm. The MI-MTVE algorithm was then applied to arrive at a final model. For the complete data and complete-case analyses the algorithm was applied using the single complete-data or complete-case data set. A p-value cut-off of 0.01 was used in the model selection. In the MI analyses we used 20 imputations. 

The results are shown in Table \ref{table:rotterdam} and Figure \ref{fig:rotterdam}. In tests of the proportional hazards assumption for individual covariates, all methods identified strong evidence for a TVE for all variables except hormone therapy (Table \ref{table:rotterdam}(a)). However, the functional form for the TVE which gave the smallest p-value in the test differed across methods. Covariate $\log (\mathrm{pgr}+1)$ was selected to the final model with a TVE under all methods, and tumour size 1 was selected to the final model with a TVE in all analyses except the complete-case analysis. The enodes covariate was identified as having a TVE in the final model using the MI analyses, but not the complete-data or complete-case analyses. Age was identified as having a TVE in the final model using the complete-case analysis but not the other methods. The MI methods gave similar estimated TVE forms for $\log (\mathrm{pgr}+1)$, tumour size 1 and enodes (Figure \ref{fig:rotterdam}). Using MI-TVE-SMC gave very wide confidence bounds for the $\log (\mathrm{pgr}+1)$ estimates, while the other MI methods performed better, giving narrower confidence bounds than under the complete-case analysis. The figure showing results for tumour size 1, which was identified to have a TVE under all methods apart from the complete-case analysis, illustrates that ignoring the missing data could result in qualitatively different conclusions about the nature of the association of this variable with the outcome. For covariates without TVEs, all methods gave similar estimated hazard ratios, while the standard errors from the MI analyses were smaller than those from the complete-case analysis, illustrating the loss of efficiency from the complete-case analysis (Table \ref{table:rotterdam}(b)). 


\section{Discussion} \label{sec:discussion}

In this article we have introduced two multiple imputation methods allowing for time-varying effects (TVEs) to be included in Cox regression models. In the absence of TVEs, the methods of White and Royston \cite{White:2009} (MI-Approx) and Bartlett et al \cite{Bartlett:2015} (MI-SMC) can be used. MI-Approx is conceptually simpler, more convenient to code and faster to run, while MI-SMC method has better statistical properties. Our two proposals are extensions of these methods. The methods were described for a general functional form for TVEs. Researchers use different approaches to modelling TVEs. The correct functional form for a TVE is typically not known in advance, and so it is desirable to allow a flexible form. We therefore focused on a situation in which TVEs are modelled using restricted cubic splines. In some studies it may be desirable for the TVE to be a simply step function and we provided details on this in the Supplementary Materials (Section S3). In simulation studies, we used imputation model assuming a 5-knot restricted cubic spline for the TVE. For binary covariates with missing data, both of our proposed methods performed well. However, the performance of the approximate method (MI-TVE-Approx) was slightly disappointing for continuous covariates when the effect size is large, though it still outperformed complete case analysis in all but scenario 2, and the observed bias was found to be smaller when the proportion with missing data is lower. The SMC method (MI-TVE-SMC) performed well across all scenarios, minimising bias, retaining the size of tests for non-proportional hazards and maximising power compared to other methods across all scenarios. Our results showed that ignoring TVEs in the imputation model will result in incorrect type I errors in the test for non-proportional hazards when the null hypothesis of proportional hazards is true, and in a large loss of power to detect a TVE when one exists. 

In practice, TVEs will often appear in the context of model building, including tests of the proportional hazards assumption. We therefore proposed the MI-MTVE model selection algorithm, an adaptation the `MFPT' algorithm of Sauerbrei et al. \cite{Sauerbrei:2007}, for such settings. We applied our proposed methods to the analysis of the Rotterdam breast cancer study data, followed by the MI-MTVE model selection algorithm. The methods led to different results, demonstrating that the choice of method will impact on substantive conclusions. Our algorithm draws on earlier work by Wood et al. \cite{Wood:2008} on variable selection methods using multiply imputed data. There is a sizeable literature on model selection incorporating estimation of TVEs, without including treatment of missing data. Berger et al (2003) \cite{Berger:2003} proposed the use of fractional polynomials \cite{Royston:1994} to select parsimonious forms for TVEs in Cox regression. Sauerbrei et al (2007) \cite{Sauerbrei:2007} proposed a model selection algorithm for use in Cox regression in which both the functional form for continuous covariates and the functional form for TVEs of covariates are modelled using fractional polynomials (the MFPT algorithm). See also Royston and Sauerbrei (2008) \cite{Royston:2008} (Chapter 11). Abrahamowicz et al (2007) \cite{Abrahamowicz:2007} and Wynant and Abrahamowicz (2014) (\cite{Wynant:2014} also described methods for joint estimation of time-varying and non-linear effects based on splines. Areas for further work include the extension of the methods proposed in this paper to a setting in which TVEs are modelled using fractional polynomials, and to allow selection of functional forms for continuous variables and covariate interactions. This will build on the work Morris et al (2015) \cite{Morris:2015} on how to incorporate MI into a fractional polynomial model building procedure for explanatory variables. The MI-TVE-SMC approach is particularly suitable for extensions involving transformed covariates. Finally, further work is needed to investigate the validity of inferences following data-dependent model selection processes in the missing data context. 

In the Supplementary Materials we provide example R code which can be used to implement the proposed imputation models. MI-TVE-Approx is straightforward to apply in standard software, and although we provide example R code, this method can also be easily applied in Stata (\texttt{mi impute}) or SAS (\texttt{PROC MI}), for example. MI-SMC (not incorporating TVEs) can be applied using the \texttt{smcfcs} package in R and Stata \cite{BartlettMorris:2015}. We have also provided an adaptation of this code for implementing MI-TVE-SMC in the setting with two covariates with TVEs, as in the simulation studies, and work is underway to make a more general version available. 

There are of course limitations to this work. In particular, in some settings complete case analysis will be unbiased and MI biased. It follows that our methods are only applicable to settings in which MI is judged to be the best approach. When using our proposed methods, other forms of mis-specification of the imputation model could result in bias and this should be borne in mind, as in any MI analysis, especially for partially missing variables which are continuous for which the normality assumption may not hold. Further work is also need to investigate the performance of different approaches to model selection in this context. The general results given for MI-TVE-Approx and MI-TVE-SMC assumed that any censoring occurs independently of covariates with missing data. In MI-TVE-Approx censoring depending on covariates with missing data can be accommodated by adding a further term, $\widehat{H}_{C}(T)$, into the imputation model, which denotes the Nelson-Aalen estimate of the cumulative hazard for the censoring \cite{Borgan:2015}. MI-SMC has been extended to allow competing risks \cite{BartlettTaylor:2016} and this can be used to handle dependence of right-censoring on variables with missing data by modelling the censoring as a competing event. MI-TVE-SMC can be extended in the same way. In both cases, it is assumed that the association between covariates and the hazard for censoring is not time-varying. Event times are also commonly subject to left-truncation. Using MI-TVE-Approx, it can be shown that left-truncation can be accommodated by replacing $\widehat{H}(T)$ by $\widehat{H}(T)-\widehat{H}(T_{L})$ (and $\widehat{H}^{(1)}(T)$ by $\widehat{H}^{(1)}(T)-\widehat{H}^{(1)}(T_{L})$) in the imputation model, and also $\widehat{H}_{C}(T)$ by $\widehat{H}_{C}(T)-\widehat{H}_{C}(T_{L})$ if the censoring is suspected to depend on partially missing covariates. MI-TVE-SMC can also be extended to accommodate left-truncation, however further work is needed on this topic, including to implement the methods in the software.
 
We have focused on estimation of TVEs by modelling these in the Cox regression model. There exist other methods for estimating and testing for TVEs in Cox regression. Schoenfeld residual plots can be used to visually assess the proportional hazards assumption \cite{Schoenfeld:1982} and smoothed residuals can be used to estimate TVEs \cite{Grambsch:1994,Winnett:2003}. Scheike and Martinussen (2004) \cite{Scheike:2004} outlined tests for proportional hazards based on an iterative procedure to estimate cumulative regression coefficients, which does not require specification of the functional form for time-varying effects. Ng'andu (1997) \cite{Ngandu:1997} summarized and compared several tests for the proportional hazards assumption.  Other methods for estimating TVEs include those based on a kernel-weighted local partial likelihood \cite{Tian:2005} and penalized partial likelihoods \cite{Zucker:1990,Verweij:1995,Yan:2012}. Buchholz and Sauerbrei (2011) \cite{Buchholz:2011} proposed a  measure for use in choosing between different models for the time-varying effect to discover which is closest to the true shape. 
 
In summary, for settings in which MI is judged to be appropriate and TVEs are a feature of the analysis, the approaches we have described should be used. Where computational time is not too large, the MI-TVE-SMC approach is recommended, though MI-TVE-Approx should also perform well if all covariates with missingness are binary or if effect sizes are small. Ignoring TVEs in the imputation may result in biased estimates and misleading conclusions.

\textbf{Acknowledgements} The authors are grateful to Dr Ian White (MRC Clinical Trials Unit at UCL, UK), Professor Mike Kenward (Department of Medical Statistics, London School of Hygiene and Tropical Medicine, UK) and Dr Jonathan Bartlett (Statistical Innovation Group, AstraZeneca, UK) for comments on this work and to Professor Patrick Royston (MRC Clinical Trials Unit at UCL, UK) for advice on the example.

Ruth Keogh is funded by a Medical Research Council Methodology Fellowship (MR/M014827/1).

\clearpage

\begin{figure}\caption{Curve-wise estimates of TVEs for covariate $X_{1}$ in the setting with binary covariates $X_{1}$ and $X_{2}$. The thick dotted black line indicates the true curve.}
\begin{center}
\includegraphics[width=.8\textwidth]{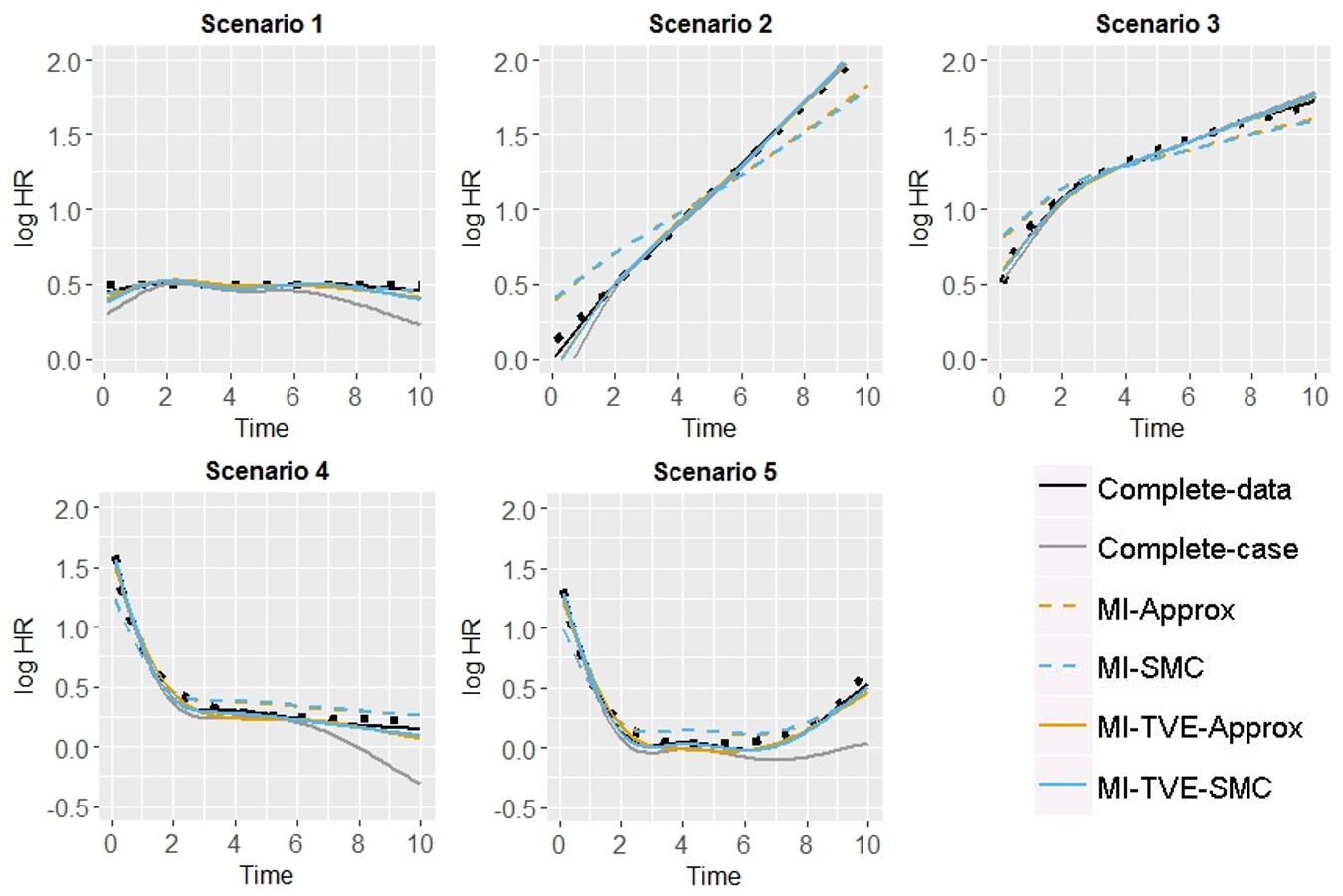}
\end{center}
\end{figure}

\begin{figure}\caption{Curve-wise estimates of TVEs for covariate $X_{1}$ in the setting with continuous covariates $X_{1}$ and $X_{2}$. The thick dotted black line indicates the true curve.}
\begin{center}
\includegraphics[width=.8\textwidth]{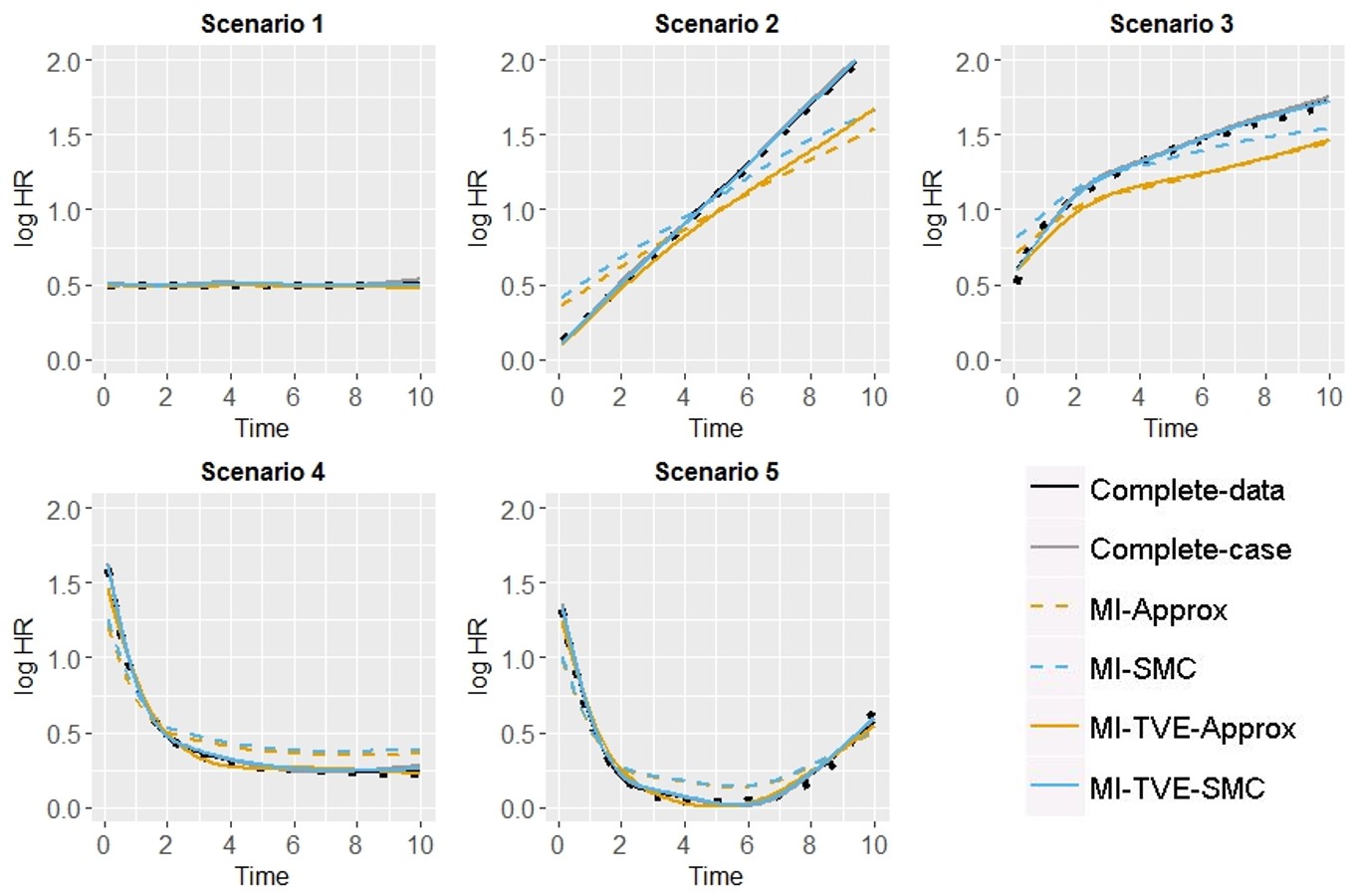}
\end{center}
\end{figure}

\begin{figure}\caption{Bias in the estimated TVE curve at three time points for covariate $X_{1}$ in the setting with binary covariates $X_{1}$ (black) and $X_{2}$ (grey). The point indicates the bias and the bar indicates the 95\% confidence interval.}
\begin{center}
\includegraphics[scale=1]{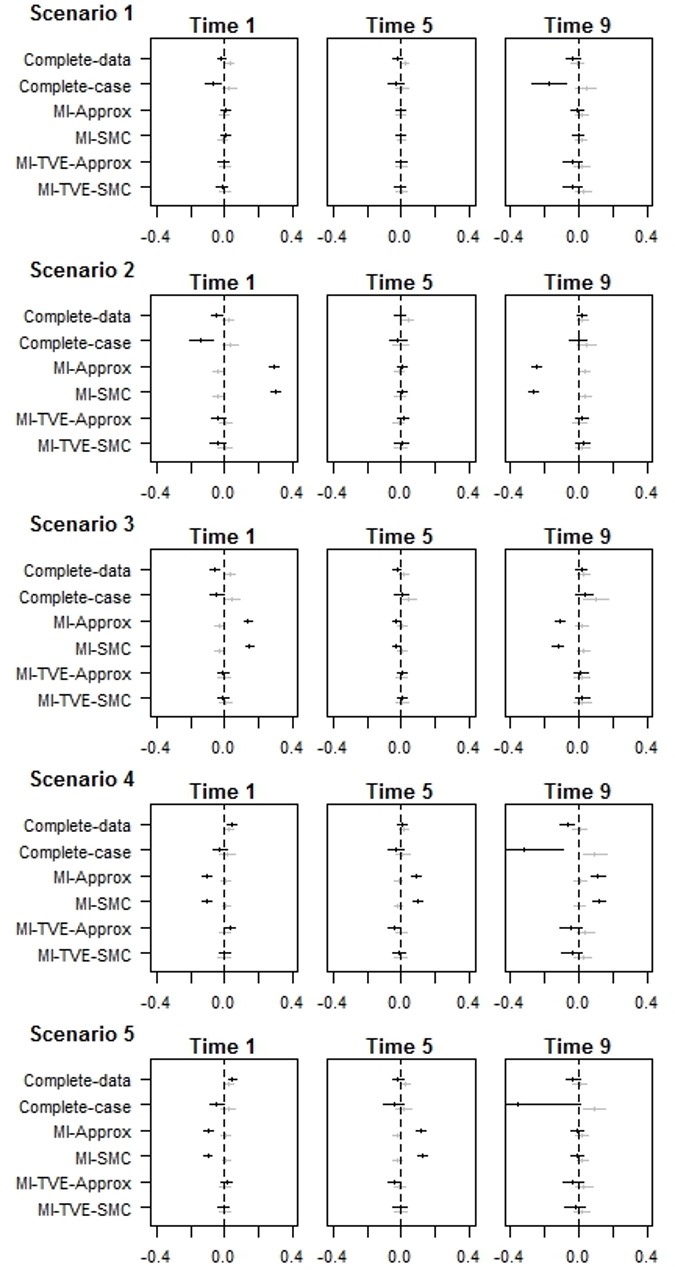}
\end{center}
\end{figure}

\begin{figure}\caption{Bias in the estimated TVE curve at three time points for covariate $X_{1}$ in the setting with continuous covariates $X_{1}$ (black) and $X_{2}$ (grey). The point indicates the bias and the bar indicates the 95\% confidence interval.}
\begin{center}
\includegraphics[scale=1]{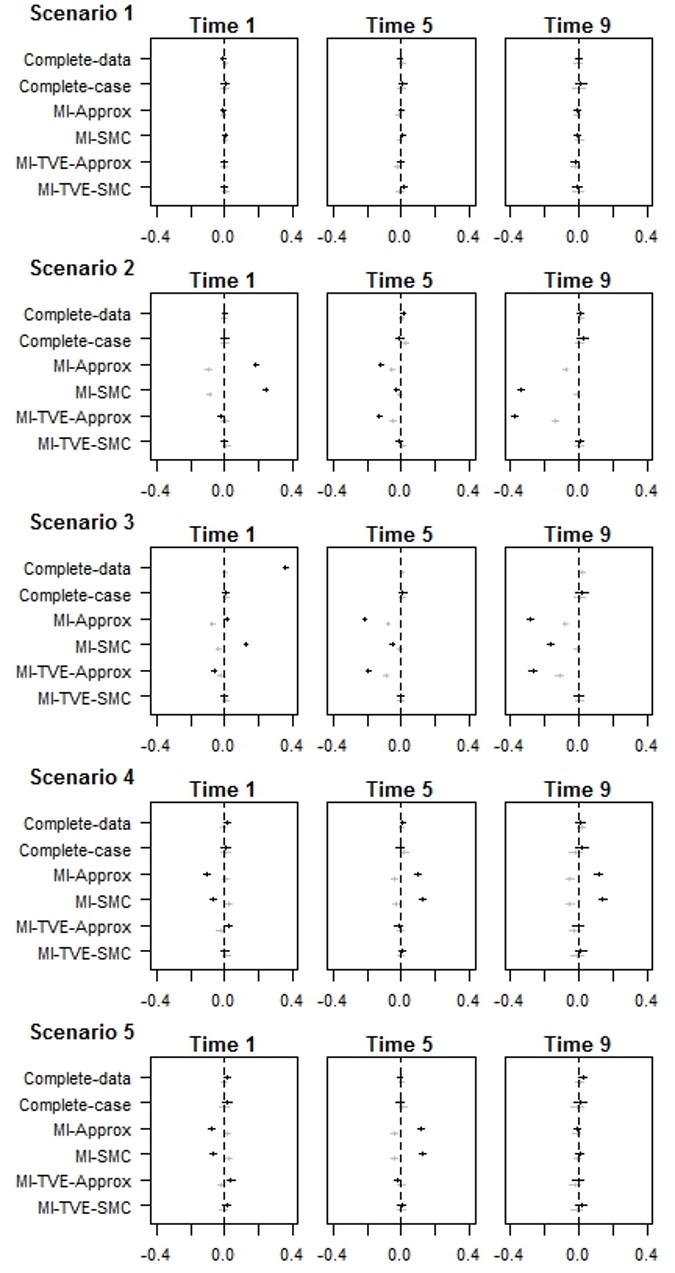}
\end{center}
\end{figure}

\begin{table}[ht]
	\caption{Percentage of simulations in which the null hypotheses of proportional hazards for $X_{1}$ and $X_{2}$ were rejected using joint Wald tests for binary (left columns) and continuous (right columns) exposures. For a given rejection percentage $\pi$, the Monte Carlo SE is $\sqrt{\tfrac{\pi(100-\pi)}{500}\times \tfrac{1}{100}}$}
	\centering
	\begin{tabular}{l|rl|rl|rl|rl|rl}
		\hline
        & \multicolumn{2}{c}{Scenario 1} & \multicolumn{2}{c}{Scenario 2} & \multicolumn{2}{c}{Scenario 3} & \multicolumn{2}{c}{Scenario 4} & \multicolumn{2}{c}{Scenario 5} \\
		& $X_1$ & $X_2$ & $X_1$ & $X_2$ & $X_1$ & $X_2$ & $X_1$ & $X_2$ & $X_1$ & $X_2$ \\ 
		\hline
		\textit{Binary $X1,X2$} & & & & & & & & & & \\
		Complete data &  3 &  3 & 89 &  3 & 33 &  3 & 56 &  6 & 45 &  4 \\
		Complete case &  2 &  3 & 42 &  3 & 14 &  2 & 17 &  2 & 14 &  3 \\
		MI-Approx     &  0 &  0 & 21 &  0 &  3 &  0 &  4 &  0 &  2 &  0 \\
		MI-SMC        &  0 &  0 & 17 &  0 &  2 &  0 &  3 &  0 &  1 &  0 \\
		MI-TVE-Approx &  2 &  3 & 67 &  3 & 16 &  2 & 27 &  2 & 21 &  2 \\
		MI-TVE-SMC    &  3 &  4 & 68 &  6 & 24 &  6 & 34 &  5 & 27 &  5 \\
		\hline
		\textit{Continuous $X1,X2$} & & & & & & & & & & \\
		Complete data &  7 &  5 & 100&  3 & 79 &  5 & 99 &  3 & 96 &  4 \\
		Complete case &  6 &  6 & 94 &  5 & 42 &  4 & 77 &  4 & 60 &  5 \\
		MI-Approx     &  0 &  0 & 68 &  0 &  6 &  0 & 43 &  0 & 27 &  0 \\
		MI-SMC        &  0 &  0 & 82 &  0 & 12 &  0 & 46 &  0 & 29 &  0 \\
		MI-TVE-Approx &  5 &  5 & 90 &  3 & 26 &  4 & 86 &  4 & 71 &  4 \\
		MI-TVE-SMC    & 10 &  9 & 99 &  8 & 57 &  6 & 89 &  6 & 78 &  8 \\
		\hline
        \end{tabular}
\end{table}

  \begin{figure}
\caption{Results from additional simulations in which the probability of missingness in $X_{1}$ and $X_{2}$ depends on $D$. Results are from scenario 4 in the situation with binary covariates. The upper-left plot shows the curve-wise estimates of TVEs for covariate $X_{1}$.  The thick dotted black line indicates the true curve. The upper-right table shows the percentage of simulations in which the null hypotheses of proportional hazards for $X_{1}$ and $X_{2}$ were rejected using joint Wald tests. The lower plot shows the bias in the estimated TVE curve at three time points for $X_{1}$ (in black) and $X_{2}$ (grey)}\label{fig:sim.MARd}
  \begin{minipage}[b]{0.5\textwidth}
    \centering
\includegraphics[scale=0.65]{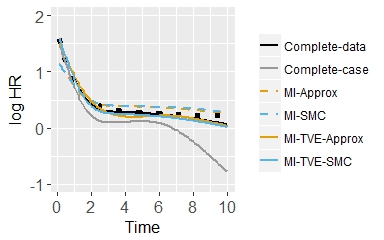}
  \end{minipage}
  \begin{minipage}[b]{0.35\textwidth}
    \centering
    \begin{tabular}{rrr}
  \hline
 & X1 & X2 \\ 
  \hline
Complete data & 58 & 5 \\ 
  Complete case & 14 & 2 \\ 
  MI-Approx & 2 & 0 \\ 
  MI-SMC & 1 & 0 \\ 
  MI-TVE-Approx & 30 & 4 \\ 
  MI-TVE-SMC & 38 & 0 \\ 
   \hline
\end{tabular}
  \vspace{1.2cm}
    \end{minipage}
  \includegraphics[scale=0.8]{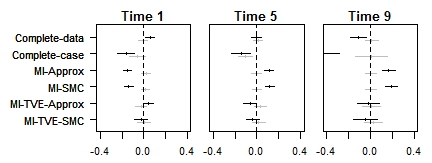}  
  \end{figure}

\begin{figure}
\caption{Results from additional simulations in which 50\% of individuals had the event. Results are from scenario 4 in the situation with binary covariates. The upper-left plot shows the curve-wise estimates of TVEs for covariate $X_{1}$.  The thick dotted black line indicates the true curve. The upper-right table shows the percentage of simulations in which the null hypotheses of proportional hazards for $X_{1}$ and $X_{2}$ were rejected using joint Wald tests. The lower plot shows the bias in the estimated TVE curve at three time points for $X_{1}$ (in black) and $X_{2}$ (grey)}\label{fig:sim.50pcevent}
  \begin{minipage}[b]{0.5\textwidth}
    \centering
\includegraphics[scale=0.65]{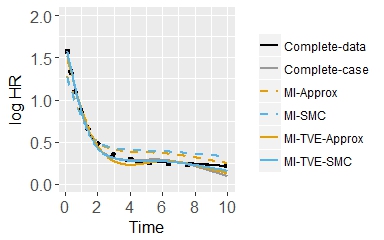}
  \end{minipage}
  \begin{minipage}[b]{0.35\textwidth}
    \centering
    \begin{tabular}{rrr}
  \hline
 & X1 & X2 \\ 
  \hline
Complete data & 100 & 3 \\ 
  Complete case & 96 & 5 \\ 
  MI-Approx & 90 & 0 \\ 
  MI-SMC & 86 & 0 \\ 
  MI-TVE-Approx & 99 & 3 \\ 
  MI-TVE-SMC & 100 & 9 \\ 
   \hline
\end{tabular}
  \vspace{1.2cm}
    \end{minipage}
  \includegraphics[scale=0.8]{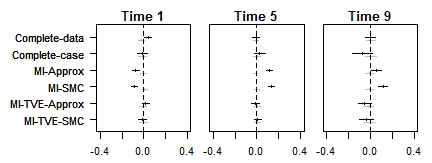}  
  \end{figure}
  
  \begin{figure}
\caption{Results from additional simulations in which the proportion of individuals missing $X_{1}$ or $X_{2}$ was reduced to 20\%. Results are from scenario 2 in the situation with continuous covariates. The upper-left plot shows the curve-wise estimates of TVEs for covariate $X_{1}$. The upper-right table shows the percentage of simulations in which the null hypotheses of proportional hazards for $X_{1}$ and $X_{2}$ were rejected using joint Wald tests. The lower plot shows the bias in the estimated TVE curve at three time points for $X_{1}$ (in black) and $X_{2}$ (grey)}\label{fig:sim.10pcmiss}
  \begin{minipage}[b]{0.5\textwidth}
    \centering
\includegraphics[scale=0.65]{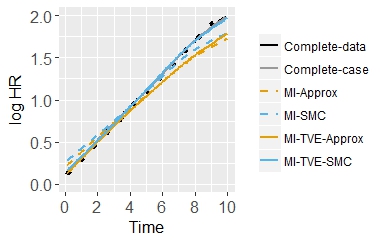}
  \end{minipage}
  \begin{minipage}[b]{0.35\textwidth}
    \centering
    \begin{tabular}{rrr}
  \hline
 & X1 & X2 \\ 
  \hline
Complete data & 100 & 5 \\ 
  Complete case & 100 & 3 \\ 
  MI-Approx & 99 & 0 \\ 
  MI-SMC & 100 & 1 \\ 
  MI-TVE-Approx & 99 & 2 \\ 
  MI-TVE-SMC & 100 & 5 \\ 
   \hline
\end{tabular}
  \vspace{1.2cm}
    \end{minipage}
  \includegraphics[scale=0.8]{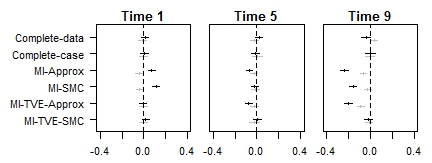}  
  \end{figure}

\begin{table} \label{table:rotterdam}
\caption{Results from the Rotterdam Breast Cancer Study.}\label{table:rotterdam}
\begin{subtable}[t]{1\linewidth}
\caption{p-values from joint Wald tests of the null hypothesis of no TVEs for each covariate, based on the model which gave the smallest p-value, and the form of that model (`p (form*)'). $^\star$`lin' denotes a TVE of linear form $f_{X}(t;\beta_{X})=\beta_{0}+\beta_{1}t$. `k3', `k4', `k5' denote restricted cubic spline forms for the TVE with 3, 4 and 5 knots.}
{\footnotesize
\begin{center}
\begin{tabular}{lrrrrrr}
\hline
Covariate & Complete-data & Complete-case& MI-Approx & MI-SMC & MI-TVE-Approx & MI-TVE-SMC \\
\hline
Age				& 0.011 (k4)  & $<0.001$ (k4)  & $<0.001$ (k4)& $<0.001$ (k4)  & $<0.001$ (k4)  & $<0.001$ (k4) \\
Size 1			& $<0.001$ (lin) & $<0.001$ (lin) & $<0.001$ (k3)& $<0.001$ (k3)  & $<0.001$ (k3)  & $<0.001$ (k3) \\
Size 2			& 0.003 (lin) & 0.002 (lin) & 0.003 (lin)& 0.003 (lin) & 0.004 (lin) & 0.003 (lin) \\
Grade			& 0.054 (lin) & 0.195 (lin) & $<0.001$ (k4)& $<0.001$ (k4)  & $<0.001$ (k4)  & $<0.001$ (k4) \\
enodes			& $<0.001$ (lin) & $<0.001$ (lin) & $<0.001$ (k4)& $<0.001$ (k4)  & $<0.001$ (k4)  & $<0.001$ (k4) \\
Hormone therapy	& 0.122 (k5)  & 0.272 (k5)  & 0.622 (lin)& 0.522 (lin) & 0.581 (lin) & 0.586 (lin) \\
Chemotherapy	& 0.007 (k4)  & 0.006 (k3)  & 0.004 (k3)& 0.004 (k3)  & 0.004 (k3)  & 0.004 (k3) \\
log(pgr+1)		& $<0.001$ (k3)  & $<0.001$ (k3)  & $<0.001$ (k4)& $<0.001$ (k4)  & $<0.001$ (k4)  & $<0.001$ (k3)\\
\hline
\end{tabular}
\end{center}}
\end{subtable}
\begin{subtable}[t]{1\linewidth}
\caption{Estimated log hazard ratios and standard errors (`Est(SE)') from the final model selected using the MI-MTVE algorithm, for covariates with no TVE. Log hazard ratios for covariates with TVEs are shown graphically in Figure \ref{fig:rotterdam}, and the corresponding covariates are labelled 'TVE' in the table.}
{\footnotesize
\begin{center}
\begin{tabular}{lllllll}
\hline
Covariate & Complete-data & Complete-case& MI-Approx & MI-SMC & MI-TVE-Approx & MI-TVE-SMC \\
\hline
Age				&-0.013 (0.002)&TVE (k4)&-0.013 (0.002)&-0.013 (0.002)&-0.013 (0.002) &-0.013 (0.002)\\
Size 1			&TVE (lin)&0.250 (0.066)&TVE (lin)&TVE (lin)&TVE (lin)&TVE (lin) \\
Size 2			&0.151 (0.081)&0.160 (0.089)&0.132 (0.081) &0.133 (0.081)&0.137 (0.082)&0.129 (0.082) \\
Grade			&0.375 (0.065)&0.367 (0.072)&0.364 (0.066)&0.367 (0.067)&0.356 (0.068) & 0.371 (0.067)\\
enodes			&-1.697 (0.084)&TVE (lin)&TVE (k4)&TVE (k4)&TVE (k4)&TVE (k4) \\
Hormone therapy	&-0.413 (0.085)&-0.334 (0.099)&-0.441 (0.090)&-0.430 (0.088)&-0.432 (0.091)&-0.442 (0.089) \\
Chemotherapy	&-0.447 (0.073)&-0.423 (0.077)&-0.428 (0.074)&-0.451 (0.074)&-0.434 (0.074) &-0.446 (0.074)\\
log(pgr+1)		&TVE (k3)&TVE (k3)&TVE (k4)&TVE (k4)&TVE (k4) &TVE (k3)\\
\hline
\end{tabular}
\end{center}}
\end{subtable}
\end{table}

\begin{figure}\caption{Results from the Rotterdam Breast Cancer Study. Plots showing estimated log hazard ratios as a function of time, for variables found to have a TVE using one or more methods. Thick lines indicate the estimates and the thin lines indicate corresponding 95\% confidence bounds.}\label{fig:rotterdam}
\begin{center}
\includegraphics[scale=0.8]{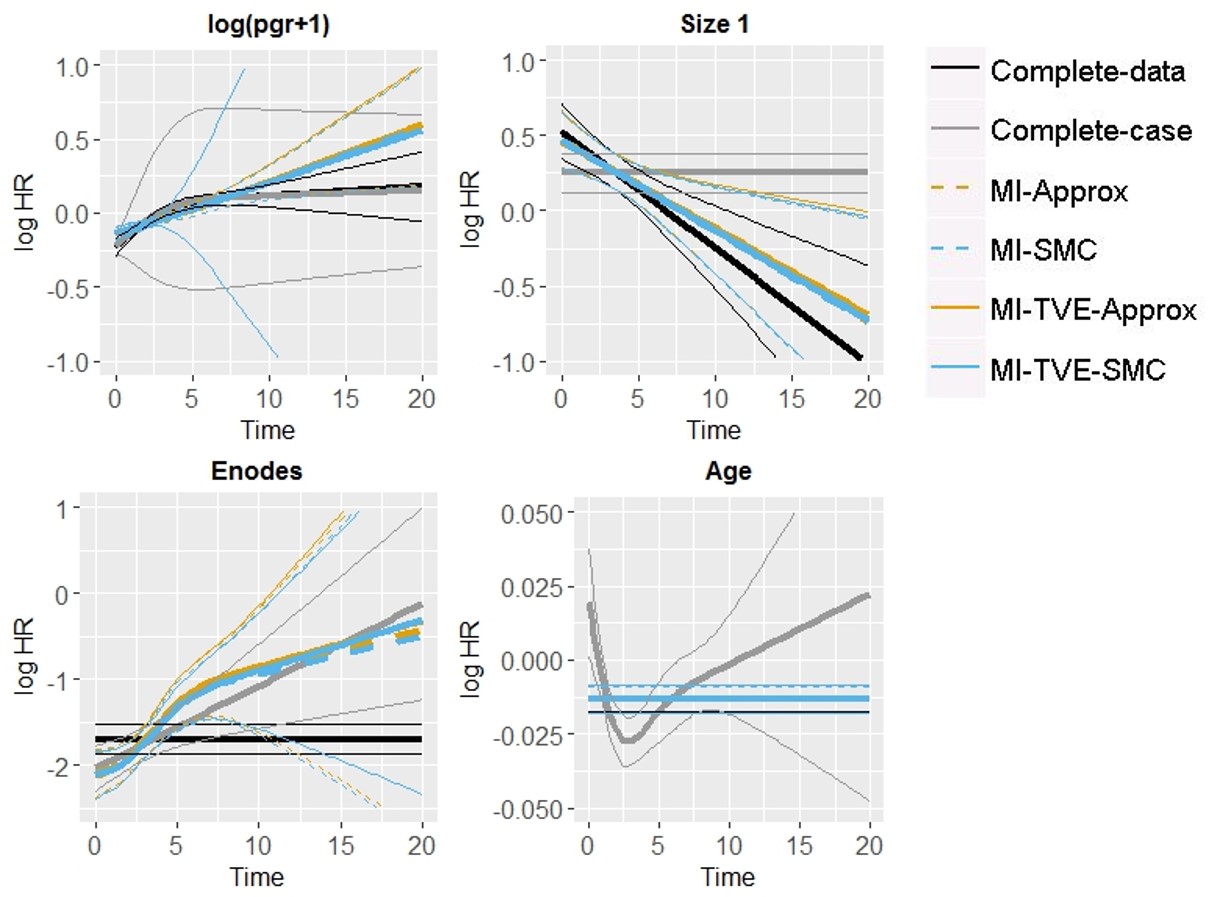}
\end{center}
\end{figure}

\clearpage

\bibliographystyle{wileyj} \bibliography{arxiv_bib}

\clearpage

\setcounter{section}{0}
\setcounter{table}{0}
\setcounter{equation}{0}
\setcounter{figure}{0}
\renewcommand{\thesection}{S\arabic{section}}
\renewcommand{\thesubsection}{S\arabic{subsection}}
\renewcommand{\thetable}{S\arabic{table}}
\renewcommand{\theequation}{S\arabic{equation}}
\renewcommand{\thefigure}{S\arabic{figure}}

\begin{center}
{\Huge\textbf{Supplementary Materials}}
\end{center}

\section{Derivation of imputation models in MI-TVE-Approx}

The focus is on a single explanatory variable $X_{1}$ with missing data and a fully observed covariate, $X_{2}$. The hazard model of interest is $h(t|X_{1},X_{2})=h_{0}(t)\exp\{f_{X_{1}}(t;\bm{\beta}_{X1})X+f_{X_{2}}(t;\bm{\beta}_{X2})X_{2}\}$. In the context of Cox regression, MI relies of obtaining draws of missing values of $X_{1}$ from its distribution given $T,D,X_{2}$. The probability density function for the conditional distribution of $X_{1}$, which we denote by $p(X_{1}|T,D,X_{2})$, can be expressed as
\begin{equation}
p(X_{1}|T,D,X_{2})=p(T,D|X_{1},X_{2})p(X_{1}|X_{2})/p(T,D|X_{2}).
\label{dist.x.tdz}
\end{equation}
The first term can be written as
\begin{equation}
p(T,D|X_{1},X_{2})=h(T|X_{1},X_{2})^{D}S(T|X_{1},X_{2})h_{C}(T|X_{1},X_{2})^{1-D}S_{C}(T|X_{1},X_{2})
\label{dist.td.xz1}
\end{equation}
where $S(.)$ is the survivor function for the event of interest, $h_{C}(t|X_{1},X_{2})$ is the hazard for censoring, and $S_{C}(.)$ is the survivor function corresponding to the censoring process. We assume for now that any censoring occurs independently of $X_{1}$ and so the third and fourth terms of (\ref{dist.td.xz1}) can be ignored in the workings which follow, since they do not involve $X_{1}$, and so can be subsumed into a constant of proportionality. Details on handling censoring which depends on $X_{1}$ are given in the Discussion section of the paper. Under the hazard model of interest we have
\begin{equation}
\begin{split}
\log p(T,D|X_{1},X_{2})=D\log h_{0}(T)+D\{f_{X1}(T,\beta_{X1})X_{1}+f_{X2}(T,\beta_{X2}) X_{2}\}\\ -\int_{0}^{T}h_{0}(u)e^{f_{X1}(u,\beta_{X1})X_{1}+f_{X2}(u,\beta_{X2}) X_{2}}\mathrm{d}u.
\end{split}
\end{equation}
It follows that 
\begin{equation}
\begin{split}
\log p(X_{1}|T,D,X_{2})=\log p(X_{1}|X_{2})+D f_{X1}(T,\beta_{X1})X_{1}\\
-\int_{0}^{T}h_{0}(u)e^{f_{X1}(u,\beta_{X1})X_{1}+f_{X2}(u,\beta_{X2}) X_{2}}\mathrm{d}u+q(T,D,X_{2})
\end{split}
\end{equation}
where $q(T,D,X_{2})$ represents terms not involving $X_{1}$.

In the situation with time varying effects the expression $\log p(X_{1}|T,D,X_{2})$ is complicated by the presence of $e^{f_{X1}(u,\beta_{X1})X_{1}+f_{X2}(u,\beta_{X2}) X_{2}}$ in the integral. For some forms for the TVE functions a closed form solution to the integral would be possible, however a more general result is desirable. A linear approximation to $e^{f_{X1}(u,\beta_{X1})X_{1}+f_{X2}(u,\beta_{X2}) X_{2}}$ is therefore used. The linear approximation is:
\begin{equation}
\begin{split}
e^{f_{X1}(u,\beta_{X1})X_{1}+f_{X2}(u,\beta_{X2})X_{2}}\approx e^{f_{X1}(\bar{u},\beta_{X1})X_{1}+f_{X2}(\bar{u},\beta_{X2})X_{2}}+(u-\bar{u})\{f^{\prime}_{X1}(\bar{u},\beta_{X1})X_{1}\\+f^{\prime}_{X2}(u,\beta_{X2})X_{2}\}e^{f_{X1}(\bar{u},\beta_{X1})X_{1}+f_{X2}(\bar{u},\beta_{X2})X_{2}}
\end{split}
\label{eq:linear.approx}
\end{equation}
where $\bar{u}$ denotes the mean of the observed event times. The approximation is expected to perform well when the TVEs are not too large, i.e. when the log hazard ratios at any given time are not too large. Higher order approximations could be considered, and in Section S3 we consider a stepwise approximation.
In the results given below, we let $A(X_{1},X_{2})={f_{X1}(\bar{u},\beta_{X1})X_{1}+f_{X2}(\bar{u},\beta_{X2})X_{2}}$, $B(X_{1},X_{2})=f^{\prime}_{X1}(\bar{u},\beta_{X1})X_{1}+f^{\prime}_{X2}(\bar{u},\beta_{X2})X_{2}$.

To proceed, it is necessary to make some assumptions about $p(X_{1}|X_{2})$. Next, we consider the situations of binary $X_{1}$ and Normally distributed $X_{1}$. 

\subsection*{S1.1 Binary $X_{1}$}

We assume a logistic model for $X_{1}$ given $X_{2}$:
\begin{equation}
\mbox{logit } p(X_{1}=1|X_{2})=\zeta_{0}+\zeta_{1} X_{2}.
\label{eq:x.z}
\end{equation}
First, suppose that $X_{2}$ is also binary. Then, using the approximation in (\ref{eq:linear.approx}), it can be shown that
\begin{equation}
\begin{split}
\mbox{logit }p(X_{1}=1|T,D,X_{2})\approx \zeta_{0}+\zeta_{1}X_{2}+D\times f_{X1}(T;\beta_{X1})\\+
H_{0}(T)\{A(0,X_{2})-A(1,X_{2})+\bar{u}A(1,X_{2})B(1,X_{2})-\bar{u}A(0,X_{2})B(0,X_{2})\}\\+H_{1}(T)\{A(0,X_{2})B(0,X_{2})-A(1,X_{2})B(1,X_{2})\}
\label{eq:binary.miapprox.1}
\end{split}
\end{equation}
where $H_{0}(T)$ denotes the cumulative baseline hazard and $H_{1}(T)=\int_{0}^{T}uh_{0}(u) du$. 

If $X_{2}$ is continuous, we use a bivariate linear approximation to $e^{f_{X1}(u,\beta_{X1})X_{1}+f_{X2}(u,\beta_{X2})X_{2}}$, about $\bar{u}$ and $\bar{X}_{2}$ (the sample mean of $X_{2}$). It can be shown that in this case
\begin{equation}
\begin{split}
\mbox{logit }p(X_{1}=1|t,d,X_{2})\approx \zeta_{0}+\zeta_{1}X_{2}+d\times f_{X1}(t;\beta_{X1})\\+
H_{0}(t)\{-A(1,\bar{X}_{2})-A(1,\bar{X}_{2})f_{X2}(\bar{u};\beta_{X2})(X_{2}-\bar{X}_{2})\\+A(0,\bar{X}_{2})+A(0,\bar{X}_{2})f_{X2}(\bar{u};\beta_{X2})(X_{2}-\bar{X}_{2})\\+ \bar{u}A(1,\bar{X}_{2})B(1,\bar{X}_{2})-\bar{u}A(0,\bar{X}_{2})B(0,\bar{X}_{2})\}\\+
H_{1}(t)\{A(0,\bar{X}_{2})B(0,\bar{X}_{2})-A(1,\bar{X}_{2})B(1,\bar{X}_{2})\}
\end{split}
\label{eq:binary.miapprox.2}
\end{equation}

It follows from the expressions in (\ref{eq:binary.miapprox.1}) and (\ref{eq:binary.miapprox.2}) that an approximate imputation model for $X_{1}$ is a logistic regression for $X_{1}$ with main effects of $X_{2}$, $H_{0}(T)$, $H_{1}(T)$, the interaction between $D$ and $f_{X1}(T)$, and interactions of $X_{2}$ with $H_{0}(T)$ and $H_{1}(T)$. If the TVE function is $f_{X1}(t;\beta_{X1})=\beta_{X01}+\beta_{X11}t$, for example, the imputation model should include $D$ and the interaction between $D$ and $T$. In the case of a restricted cubic spline with $L=5$ knots, the imputation model should include $D$ and the interaction between $D$ and $T$ and interactions of $D$ with $\left\{(T-u_{i})^{3}_{{\tiny +}}-\left(\frac{(T-u_{L-1})^{3}_{{\tiny +}}(u_{L}-u_{i})}{(u_{L}-u_{L-1})}\right)+\left(\frac{(T-u_{L})^{3}_{{\tiny +}}(u_{L-1}-u_{i})}{(u_{L}-u_{L-1})}\right)\right\}$ for $i=1,2,3$. 

In the situation without TVEs, the above results reduce to those of \cite{White:2009}. The imputation models involve the baseline cumulative hazard $H_{0}(T)$ and the additional integral term $H_{1}(T)$. When there are no TVEs, the imputation model includes only $H_{0}(T)$ and White and Royston (2009) \cite{White:2009} suggested replacing this with the Nelson-Aalen estimate of the cumulative hazard, $\widehat{H}(T)=\sum_{t\leq T}\frac{d(t)}{n(t)}$, where $d(t)$ is the number of events at time $t$ and $n(t)$ is the number of individuals at risk at time $t$. This has been found to perform at least as well as a more complex method using Breslow's estimate for $H_{0}(T)$ in simulation studies. Following similar reasoning, we propose using the Nelson-Aalen-type estimator $\widehat{H}^{(1)}(T)=\sum_{t\leq T}\frac{td(t)}{n(t)}$ in place of $H_{1}(T)$. 

\subsection*{S1.2 Continuous $X$}

To derive an imputation model for a continuous $X_{1}$ we assume that, conditionally on $X_{2}$, $X_{1}$ is normally distributed with mean $\zeta_{0}+\zeta_{1}X_{2}$ and variance $\sigma^{2}$. The derivations, which are not shown in detail here, use a quadratic (i.e. second order) trivariate approximation for $e^{f_{X1}(u,\beta_{X1})X_{1}+f_{X2}(u,\beta_{X2}) X_{2}}$ about $\bar{u}$, $\bar{X}_{1}$ and $\bar{X}_{2}$. It can be shown that an approximate imputation model for $X_{1}$ is a a linear regression of $X_{1}$ with main effects of $X_{2}$, $H_{0}(T)$, $H_{1}(T)$, the interaction between $D$ and $f_{X1}(T)$, and interactions of $X_{2}$ with $H_{0}(T)$ and $H_{1}(T)$. That is, the imputation model contains the same terms as for binary $X_{1}$ described above. AS above, we propose replacing $H_{0}(T)$ and $H_{1}(T)$ with the estimates $\widehat{H}(T)$ and $\widehat{H}^{(1)}$ respectively.

\section{Extensions to MI-TVE-Approx and MI-TVE-SMC: handling missing data in more than one covariate using full conditional specification (FCS)}

In this section we describe extensions to MI-TVE-Approx and MI-TVE-SMC for the situation with more than one covariate with missing data. Let $X=(X_{1},X_{2},\ldots,X_{p})^{\prime}$ denotes the vector of partially observed variables. The model of interest is assumed to be of the form 
$$
h(t|X)=h_{0}(t)\exp\left\{\sum_{k}f_{X_k}(t;\bm{\beta}_{Xk})X_{k}\right\}
$$
Additional fully observed covariates can be incorporated in a straightforward manner. For MI-TVE-approx, the FCS algorithm to generate a single imputed dataset is as follows.
\begin{enumerate}
	\item
	Replace the missing values in $X$ by arbitrary starting values, to create a complete data set.
	In practice, one could replace missing values of $X_k$ ($k=1, \ldots, p$) by the mean of $X_k$ among those individuals in whom $X_k$ is observed.
	Set $k=1$.
	\item
	If $X_k$ is a continuous variable, fit the imputation model 
	{\small
		$$
		X_{k}=\alpha_{0}+\alpha_{1}^{\prime}X_{-k}+\alpha_{2}^{T}D f_{Xk}(T)+\alpha_{3}\widehat{H}(T)+\alpha_{4}\widehat{H}^{(1)}(T)+\alpha_{5}^{\prime}X_{-k}\widehat{H}(T)+\alpha_{6}^{\prime}X_{-k}\widehat{H}^{(1)}(T)+\epsilon,
		$$}   
	with residual error variance $\sigma^{2}_{\epsilon}$, to the subset of individuals for whom $X_k$ is observed, using the current values of $X_{-k}$.
	If $X_k$ is a binary variable, the imputation model is the logistic regression
	\begin{equation}
	\nonumber \begin{split}
	\mbox{logit }\mbox{Pr}(X_{k}=1|T,D,X_{-k})=\alpha_{0}+\alpha_{1}^{\prime}X_{-k}+\alpha_{2}^{T}D f_{Xk}(T)+\alpha_{3}\widehat{H}(T)\\+\alpha_{4}\widehat{H}^{(1)}(T)+\alpha_{5}^{\prime}X_{-k}\widehat{H}(T)+\alpha_{6}^{\prime}X_{-k}\widehat{H}^{(1)}(T).
	\end{split}
	\end{equation}
	Take a random draw $(\alpha_0^*, \alpha_1^*, \alpha_2^*, \alpha_3^*, \alpha_4^*, \alpha_5^*, \alpha_6^*, \sigma^{2*}_{\epsilon})$ (if $X_k$ is continuous) or $(\alpha_0^*, \alpha_1^*, \alpha_2^*, \alpha_3^*, \alpha_4^*, \alpha_5^*, \alpha_6^*)$ (if $X_k$ is binary) from the approximate posterior distribution of the parameters in this model.
	\item
	If $X_k$ is continuous, then for each individual with missing $X_k$ in the original data set, replace the current value of $X_k$ with a sample from a normal distribution with mean $\alpha_{0}^{*}+\alpha_{1}^{\prime *}X_{-k}+\alpha_{2}^{\prime *}D f_{X1}(T)+\alpha_{3}^{*}\widehat{H}(T)+\alpha_{4}^{*}\widehat{H}^{(1)}(T)+\alpha_{5}^{\prime *}X_{-k}\widehat{H}(T)+\alpha_{6}^{\prime *}X_{-k}\widehat{H}^{(1)}(T)$ and variance $\sigma^{2*}_{\epsilon}$.
	If $X_k$ is binary, sample instead from a Bernoulli distribution with the same mean.
	\item
	If $k<p$, set $k=k+1$ and return to step 2.
\end{enumerate}
Repeat steps 2--4 until the sampled values of $X$ converge in distribution.
At this point, use these sampled values as the imputed values for the single imputed dataset.
Repeat the whole process $M$ times to generate $M$ imputed datasets.

For MI-TVE-SMC with $p$ partially observed variables, the algorithm to generate one imputed data set is as follows.

\begin{enumerate}
	\item
	Replace the missing missing values in $X$ with arbitrary starting values, to create a complete dataset.
	Set $k=1$.
	\item	Fit the Cox regression model of interest, including the TVEs, to the current complete data set to obtain estimates $\bm{\hat{\beta}}_{Xk}$ ($k=1,\ldots,p$) and their estimated variance $\widehat{\Sigma}$.
	Draw values $\bm{\beta}^*_{Xk}$ ($k=1,\ldots,p$) from a joint normal distribution with mean $(\bm{\hat{\beta}}_{X1},\ldots,\bm{\hat{\beta}}_{Xp})$ and variance $\widehat{\Sigma}$.
	\item Calculate Breslow's estimate, denoted $H^*_0(t)$, of the baseline cumulative hazard $H_{0}(t)$ using the parameter values $\bm{\beta}^*_{Xk}$ ($k=1,\ldots,p$) and the current imputations of $X$.
	\item Fit a regression model (e.g.\ linear or logistic, as appropriate) of $X_k$ on $X_{-k}$ to the current complete data set.
	Draw a value $\gamma_{Xk}^{*}$ from the approximate joint posterior distribution of the parameters $\gamma_{Xk}$ in this model.
	\item For each individual for whom $X_k$ is missing, (a) draw a value $X_k^*$ from the distribution $p(X_k| X_{-k};\gamma_{Xk}^{*})$ and let $X^*$ denote $X$ with $X_k$ replaced by its proposed value $X_k^*$, (b) draw a value $U$ from a uniform distribution on $[0,1]$, and (c) accept the proposal $X_k^{*}$ if
	{\small
		$$
		\left\{
		\begin{array}{ll}
		U\leq \exp \left[- \sum_{j:t_{j} \leq T} \Delta H^{(m)}_{0}(t_{j}) \exp\left\{\sum_{k}f_{Xk}\left(t_{j};\bm{\beta}^{(m)}_{Xk}\right)X_{k}^{*}\right\}  \right]&\mbox{if }D=0\\
		U\leq \Delta H^{(m)}_{0}(T)\exp\left\{1+f_{Xk}\left(T;\bm{\beta}^{(m)}_{Xk}\right)X_{k}^{*}- \sum_{j:t_{j} \leq T} \Delta H^{(m)}_{0}(t_{j})e^{f_{Xk}\left(t_{j};\bm{\beta}^{(m)}_{Xk}\right)X_{k}^{*}}\right\}&\mbox{if }D=1
		\end{array}\right.
		$$}
	
	If $X^*_k$ is not accepted, then discard it and repeat (a), (b) and (c).
	\item
	If $k<p$, let $k=k+1$ and return to step 2.
\end{enumerate}

Repeat steps 2--6 until the sampled values of $X$ converge in distribution.
At this point, use these sampled values as the imputed values for the single imputed dataset.
Repeat the whole process $M$ times to generate $M$ imputed datasets.

\section{Using a step-function form for the time-varying effect}

A simple approach to investigating the TVE of a covariate is to assume a step function form for $f_{X}(t;\bm{\beta})$, such that the hazard ratio is assumed constant within a series of time periods (see for example \cite{Gore:1984}). In the case, focusing on our situation with a partially observed covariate $X_{1}$ and a fully observed covariate $X_{2}$, the hazard function is 
\begin{equation}
h(t|X_{1},X_{2})=h_{0}(t)\exp\left\{\sum_{j=1}^{K}\beta_{X1j}I_{j}X_{1}+\sum_{j=1}^{K}\beta_{X2j}I_{j}X_{2}\right\}
\end{equation}
where there are $K$ time periods $(0,s_{1}], (s_{1},s_{2}], \ldots, (s_{K-1},s_{K}]$ and $I_{k}=I(s_{k-1}<t\leq s_{k})$ is an indicator taking value 1 if $t$ lies in the interval from $s_{k-1}$ to $s_{k}$  ($k=1,\ldots,K$) and 0 otherwise. A step function is unlikely to represent the true underlying time-varying effect and more realistic models are based on splines or other flexible functional forms, which are the main focus of the paper. However, because a step-function is sometimes used, we present some brief details here. By following similar workings as shown in Section S1, it can be shown that a suitable imputation model for $X_{1}$ is a logistic regression (for binary $X_{1}$) or linear regression (for continuous $X_{1}$) on $X_{2}$, $D I_{k}$ ($k=1,\ldots,K$), $H_{k}^{*}$ ($k=1,\ldots,K_{T}$), $H_{T}^{*}$, $X_{2} H_{k}^{*}$ ($k=1,\ldots,K_{T}$) and $X_{2} H_{T}^{*}$, where $H_{k}^{*}=\int_{s_{k-1}}^{s_{k}}h_{0}(u)\mathrm{d}u$, $H_{T}^{*}=\int_{s_{T}}^{T}h_{0}(u)\mathrm{d}u$, $K_{T}$ is the number of complete time periods which have passed prior to $T$, and $s_{T}$ is the upper limit of the last complete time period prior to $T$. We propose replacing $H_{k}^{*}$ and $H_{T}^{*}$ by their estimates
$\widehat{H}_{k}^{*}=\sum_{s_{k-1}<t\leq s_{k}}\frac{d(t)}{n(t)}$ and  $\widehat{H}_{T}^{*}=\sum_{s_{T}<t\leq T}\frac{d(t)}{n(t)}$. A feature of the imputation model for TVEs based on a step function is that we do not require a linear (or other) approximation to evaluate the integral in S4. 

\section{Simulation study: Details on missing data generation}

In the main simulation, non-monotone missing data were generated in $X_{1}$ and $X_{2}$ according to a MAR mechanism in which the probability of missingness in $X_{1}$ depends on observed values of $X_{2}$, and vice versa. To achieve this, the cohort was divided randomly into three groups of approximately equal size. In group 1, $X_{2}$ is fully observed and $X_{1}$ was set to be missing with probability $e^{0.4+0.5X_{2}}/(1+e^{0.4+0.5X_{2}})$. In group 2, $X_{1}$ is fully observed and $X_{2}$ was set to be missing with probability $e^{0.4+0.5X_{1}}/(1+e^{0.4+0.5X_{1}})$. In group 3, $X_{1}$ and $X_{2}$ were \textit{both} missing completely at random with probability 0.3. 

In additional simulations (Section 5.5), the probability of missingness in $X_{1}$ and $X_{2}$ additionally depends on the event indicator $D$. The procedure described above was modified such that in group 1 the probability of missingness in $X_{1}$ was $e^{-0.4+0.5X_{2}+0.5D+0.5X_{2}D}/(1+e^{-0.4+0.5X_{2}+0.5D+0.5X_{2}D})$, in group 2 probability of missingness in $X_{2}$ was $e^{-0.4+0.5X_{1}+0.5D+0.5X_{1}D}/(1+e^{-0.4+0.5X_{1}+0.5D+0.5X_{1}D})$, and in group 3 $X_{1}$ and $X_{2}$ were \textit{both} missing with probability $e^{-0.4+0.5D}/(1+e^{-0.4+0.5D})$. 

The values used in these missing data generation procedures were selected so that $X_{1}$ is missing for approximately 30\% of individuals and $X_{2}$ is missing for approximately 30\% of individuals, resulting in approximately 50\% of individuals missing at least one of the measurements.

In an additional simulation (Section 5.5) a lower proportion with missing data was considered. For this, in group 1, $X_{2}$ is fully observed and $X_{1}$ was set to be missing with probability $e^{-1.2+0.5X_{2}}/(1+e^{-1.2+0.5X_{2}})$. In group 2, $X_{1}$ is fully observed and $X_{2}$ was set to be missing with probability $e^{-1.2+0.5X_{1}}/(1+e^{-1.2+0.5X_{1}})$. In group 3, $X_{1}$ and $X_{2}$ were \textit{both} missing completely at random with probability 0.1.  

\section{Simulation study: Justification of number of simulated data sets}

The performance measures described in Section 5.3 were used to determine the number of repetitions under each scenario. We are primarily interested in bias, and assume that the variance of bias at any given $t$ is 0.1. Then the Monte Carlo standard error for the bias is
\begin{equation}
\textrm{MCSE} = \sqrt{\frac{\textrm{Var}}{\textrm{reps}}}.
\end{equation}
Aiming for MCSE of 0.015 on estimated bias, we require 445 repetitions, and rounded up to 500.

Secondary interest is in rejection fractions and coverage, for which the summary of a simulation run is binary. Here, for rejection fraction $\pi$ the MCSE is
\begin{equation}
\textrm{MCSE} = \sqrt{\frac{\pi(1-\pi)}{\textrm{reps}}}.
\end{equation}
The MCSE is maximised at $\pi=0.5$, for which 500 repetitions returns MCSE of 2.2\% , which we find acceptable. If tests have approximately the correct size, then at $\pi=0.05$, $\text{MCSE}=1\%$

\section{Example R code}

Example R code for implementation of the methods described in this paper can be found at: \url{https://github.com/ruthkeogh/MI-TVE}.

\clearpage

\begin{figure}\caption{Curve-wise estimates of TVEs for covariate $X_{2}$ in the setting with binary covariates $X_{1}$ and $X_{2}$. The thick dotted black line indicates the true curve. The curves are all approximately flat because $X_{2}$ always has a non-TVE.}
	\begin{center}
		\includegraphics[scale=0.6]{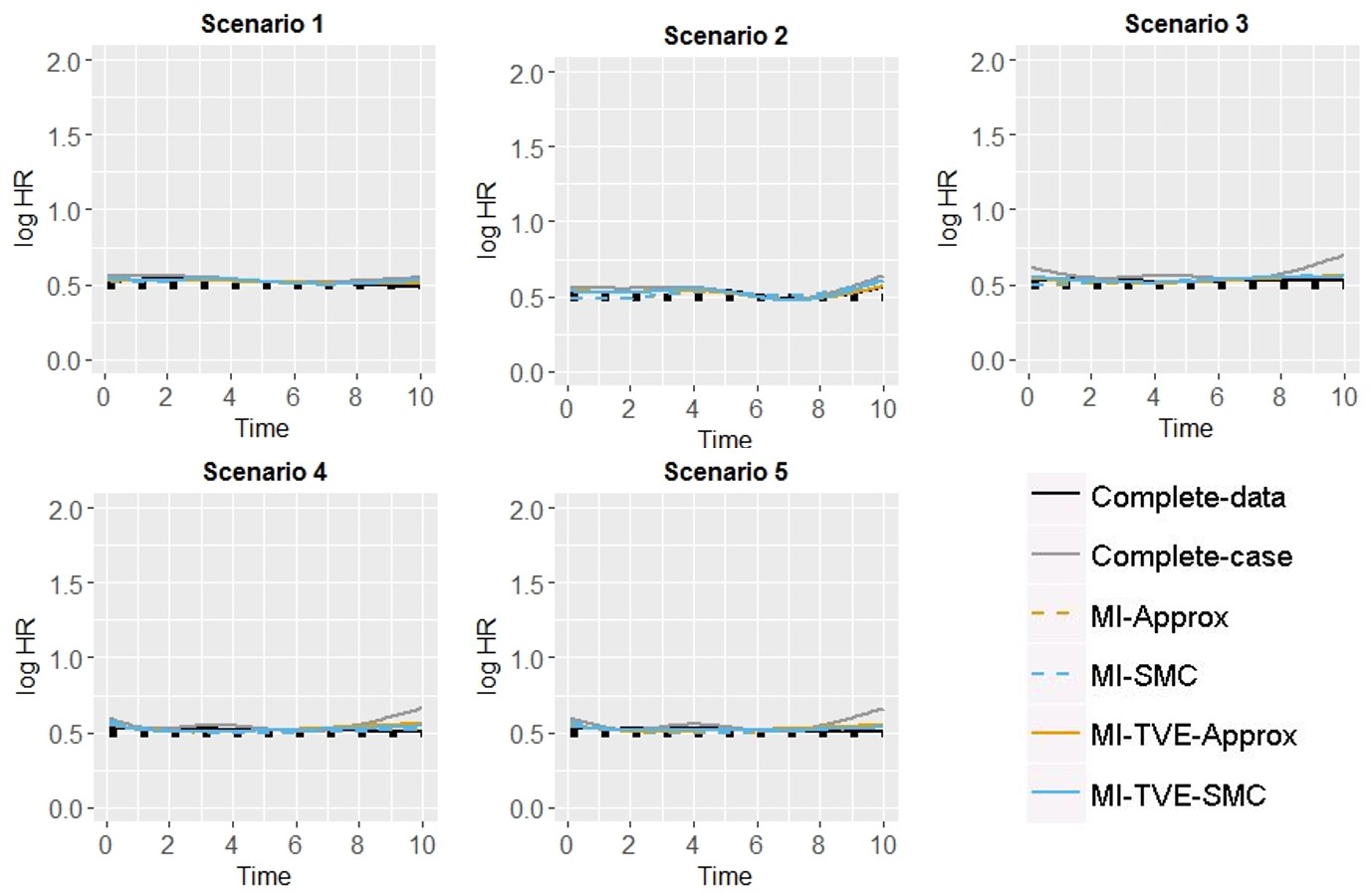}
	\end{center}
\end{figure}

\begin{figure}\caption{Curve-wise estimates of TVEs for covariate $X_{2}$ in the setting with continuous covariates $X_{1}$ and $X_{2}$. The thick dotted black line indicates the true curve. The curves are all approximately flat because $X_{2}$ always has a non-TVE.}
	\begin{center}
		\includegraphics[scale=0.6]{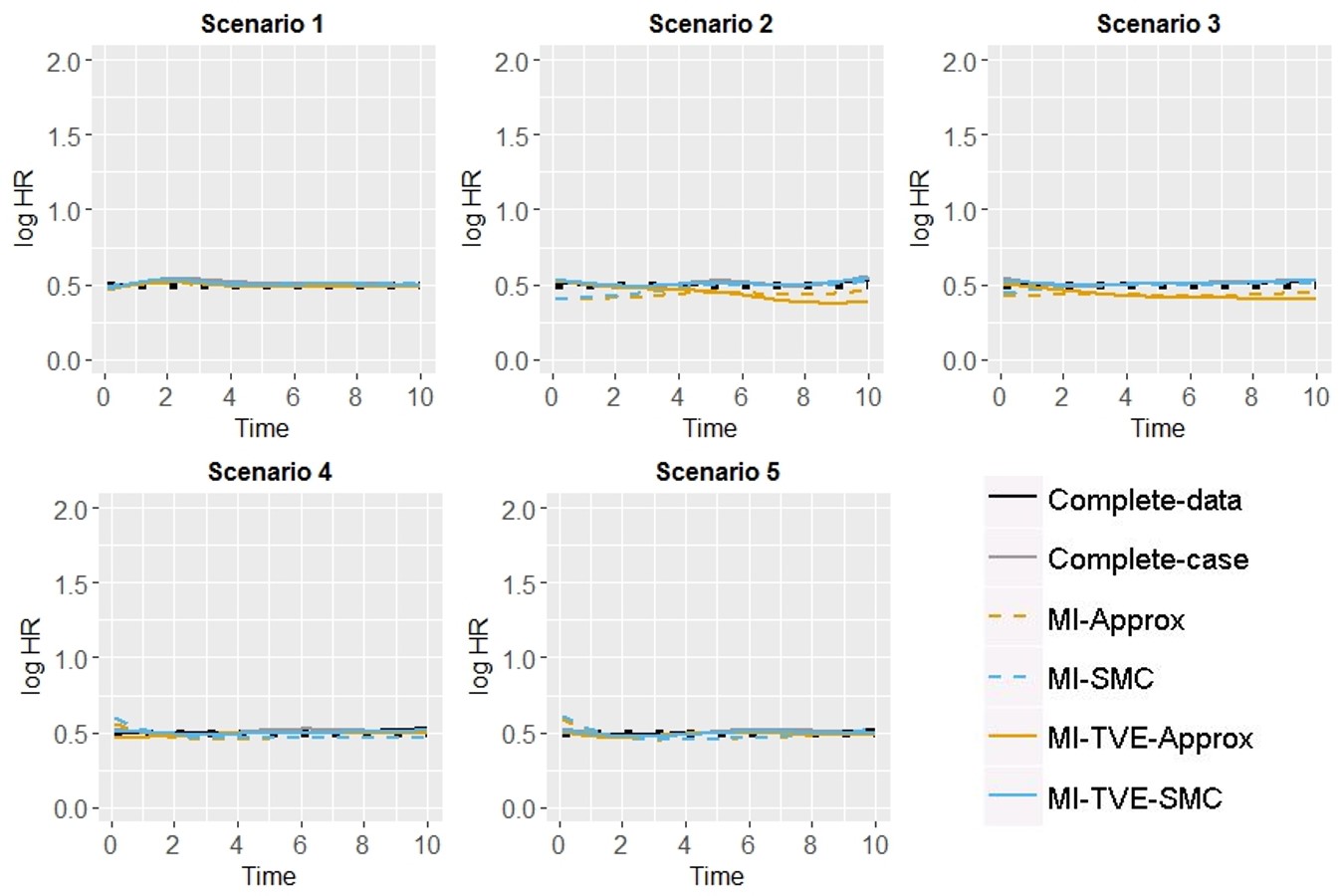}
	\end{center}
\end{figure}

\begin{table}[ht]
	\caption{ Coverage of the estimated TVE curve at three time points (1, 5, 9) for covariates $X_{1}$ and $X_{2}$ in the setting with binary covariates $X_{1}$ and $X_{2}$. }
	\centering
	\begin{tabular}{rllllll}
		\hline
		&\multicolumn{3}{c}{Covariate $X_{1}$}&\multicolumn{3}{c}{Covariate $X_{2}$}\\
		\cline{2-7}
		&1 & 5& 9 & 1& 5& 9 \\ 
		\hline
		\multicolumn{7}{l}{\textbf{Scenario 1}}\\
		Complete data & 96 & 100 & 100 & 95 & 100 & 100 \\ 
		Complete case & 98 & 100 & 100 & 95 & 100 & 100 \\ 
		MI-Approx & 99 & 100 & 100 & 98 & 100 & 100 \\ 
		MI-SMC & 100 & 100 & 100 & 97 & 100 & 100 \\ 
		MI-TVE-Approx & 96 & 100 & 100 & 95 & 100 & 100 \\ 
		MI-TVE-SMC & 97 & 100 & 100 & 95 & 100 & 100 \\ 		
		\multicolumn{7}{l}{\textbf{Scenario 2}}\\
		Complete data & 95 & 100 & 100 & 96 & 100 & 100 \\ 
		Complete case & 98 & 100 & 100 & 95 & 100 & 100 \\ 
		MI-Approx & 95 & 100 & 100 & 98 & 100 & 100 \\ 
		MI-SMC & 92 & 100 & 100 & 98 & 100 & 100 \\ 
		MI-TVE-Approx & 95 & 100 & 100 & 94 & 100 & 100 \\ 
		MI-TVE-SMC & 95 & 100 & 100 & 93 & 100 & 100 \\ 
		\multicolumn{7}{l}{\textbf{Scenario 3}}\\ 
		Complete data & 94 & 100 & 100 & 95 & 100 & 100 \\ 
		Complete case & 95 & 100 & 100 & 96 & 100 & 100 \\ 
		MI-Approx & 98 & 100 & 100 & 98 & 100 & 100 \\ 
		MI-SMC & 98 & 100 & 100 & 99 & 100 & 100 \\ 
		MI-TVE-Approx & 94 & 100 & 100 & 94 & 100 & 100 \\ 
		MI-TVE-SMC & 94 & 100 & 100 & 95 & 100 & 100 \\ 
		\multicolumn{7}{l}{\textbf{Scenario 4}}\\ 
		Complete data & 96 & 100 & 100 & 97 & 100 & 100 \\ 
		Complete case & 96 & 100 & 100 & 96 & 100 & 100 \\ 
		MI-Approx & 99 & 100 & 100 & 99 & 100 & 100 \\ 
		MI-SMC & 99 & 100 & 100 & 99 & 100 & 100 \\ 
		MI-TVE-Approx & 98 & 100 & 100 & 98 & 100 & 100 \\ 
		MI-TVE-SMC & 97 & 100 & 100 & 97 & 100 & 100 \\
		\multicolumn{7}{l}{\textbf{Scenario 5}}\\
		Complete data & 95 & 100 & 100 & 95 & 100 & 100 \\ 
		Complete case & 96 & 100 & 100 & 95 & 100 & 100 \\ 
		MI-Approx & 99 & 100 & 100 & 98 & 100 & 100 \\ 
		MI-SMC & 99 & 100 & 100 & 98 & 100 & 100 \\ 
		MI-TVE-Approx & 95 & 100 & 100 & 96 & 100 & 100 \\ 
		MI-TVE-SMC & 94 & 100 & 100 & 96 & 100 & 100 \\ 
		\hline
	\end{tabular}
\end{table}

\begin{table}[ht]
	\caption{ Coverage of the estimated TVE curve at three time points (1, 5, 9) for covariates $X_{1}$ and $X_{2}$ in the setting with continuous covariates $X_{1}$ and $X_{2}$. }
	\centering
	\begin{tabular}{rllllll}
		\hline
		&\multicolumn{3}{c}{Covariate $X_{1}$}&\multicolumn{3}{c}{Covariate $X_{2}$}\\
		\cline{2-7}
		&1 & 5& 9 & 1& 5& 9 \\ 
		\hline
		\multicolumn{7}{l}{\textbf{Scenario 1}}\\	
		Complete data & 96 & 100 & 100 & 96 & 100 & 100 \\ 
		Complete case & 95 & 100 & 100 & 97 & 100 & 100 \\ 
		MI-Approx & 99 & 100 & 100 & 99 & 100 & 100 \\ 
		MI-SMC & 99 & 100 & 100 & 98 & 100 & 100 \\ 
		MI-TVE-Approx & 97 & 100 & 100 & 97 & 100 & 100 \\ 
		MI-TVE-SMC & 97 & 100 & 100 & 96 & 100 & 100 \\ 	
		\multicolumn{7}{l}{\textbf{Scenario 2}}\\
		Complete data & 94 & 100 & 100 & 94 & 100 & 100 \\ 
		Complete case & 96 & 100 & 100 & 96 & 100 & 100 \\ 
		MI-Approx & 92 & 100 & 100 & 97 & 100 & 100 \\ 
		MI-SMC & 86 & 100 & 100 & 98 & 100 & 100 \\ 
		MI-TVE-Approx & 94 & 100 & 100 & 94 & 100 & 100 \\ 
		MI-TVE-SMC & 95 & 100 & 100 & 95 & 100 & 100 \\ 
		\multicolumn{7}{l}{\textbf{Scenario 3}}\\
		Complete data & 94 & 100 & 100 & 96 & 100 & 100 \\ 
		Complete case & 95 & 100 & 100 & 95 & 100 & 100 \\ 
		MI-Approx & 99 & 100 & 100 & 98 & 100 & 100 \\ 
		MI-SMC & 97 & 100 & 100 & 99 & 100 & 100 \\ 
		MI-TVE-Approx & 92 & 100 & 100 & 95 & 100 & 100 \\ 
		MI-TVE-SMC & 95 & 100 & 100 & 96 & 100 & 100 \\ 
		\multicolumn{7}{l}{\textbf{Scenario 4}}\\
		Complete data & 100 & 100 & 100 & 100 & 100 & 100 \\ 
		Complete case & 100 & 100 & 100 & 100 & 100 & 100 \\ 
		MI-Approx & 100 & 100 & 100 & 100 & 100 & 100 \\ 
		MI-SMC & 100 & 100 & 100 & 100 & 100 & 100 \\ 
		MI-TVE-Approx & 100 & 100 & 100 & 100 & 100 & 100 \\ 
		MI-TVE-SMC & 99 & 100 & 100 & 100 & 100 & 100 \\ 
		\multicolumn{7}{l}{\textbf{Scenario 5}}\\
		Complete data & 98 & 100 & 100 & 99 & 100 & 100 \\ 
		Complete case & 99 & 100 & 100 & 98 & 100 & 100 \\ 
		MI-Approx & 100 & 100 & 100 & 100 & 100 & 100 \\ 
		MI-SMC & 100 & 100 & 100 & 100 & 100 & 100 \\ 
		MI-TVE-Approx & 99 & 100 & 100 & 100 & 100 & 100 \\ 
		MI-TVE-SMC & 98 & 100 & 100 & 99 & 100 & 100 \\ 
		\hline
	\end{tabular}
\end{table}

\end{document}